\documentclass[journal=jctcce,manuscript=article]{achemso}

\usepackage{bm}
\usepackage{booktabs}
\usepackage{dcolumn}
\usepackage{graphicx}
\usepackage{float}

\usepackage{amsmath}
\usepackage{amssymb}
\usepackage{amsfonts}
\usepackage{braket}
\usepackage{mathtools}

\newcommand{\half}{\frac{1}{2}}
\newcommand{\bfr}{\mathbf{r}}
\newcommand{\rone}{\mathbf{r}_1}
\newcommand{\rtwo}{\mathbf{r}_2}
\newcommand{\rthr}{\mathbf{r}_3}

\title{Explicitly correlated electronic structure calculations with transcorrelated matrix product operators}

\author{Alberto Baiardi}
\affiliation{ETH Z\"urich, Laboratory of Physical Chemistry, Vladimir-Prelog-Weg 2, 8093 Z\"urich, Switzerland}

\author{Micha{\l} Lesiuk$^\ddagger$}
\affiliation{ETH Z\"urich, Laboratory of Physical Chemistry, Vladimir-Prelog-Weg 2, 8093 Z\"urich, Switzerland}
\affiliation{Faculty of Chemistry, University of Warsaw, Pasteura 1, 02-093 Warsaw, Poland}

\author{Markus Reiher}
\email{markus.reiher@phys.chem.ethz.ch}
\affiliation{ETH Z\"urich, Laboratory of Physical Chemistry, Vladimir-Prelog-Weg 2, 8093 Z\"urich, Switzerland}

\date{May 12, 2022}

\begin{document}

\begin{abstract}
\footnotetext{The authors AB and ML contributed equally to this work.}

In this work, we present the first implementation of the transcorrelated electronic Hamiltonian in an optimization procedure for
matrix product states by the density matrix renormalization group (DMRG) algorithm. 
In the transcorrelation ansatz, the electronic Hamiltonian is similarity-transformed with a Jastrow factor to 
describe the cusp in the wave function at electron-electron coalescence.
As a result, the wave function is easier to approximate accurately with the conventional expansion in terms 
of one-particle basis functions and Slater determinants.
The transcorrelated Hamiltonian in first quantization comprises up to three-body interactions, which we deal with
in the standard way
by applying robust density fitting to two- and three-body integrals entering the second-quantized representation of this Hamiltonian.
The lack of hermiticity of the transcorrelated Hamiltonian is taken care of along the lines of the first work 
on transcorrelated DMRG [J. Chem. Phys. 153, 164115 (2020)] 
by encoding it as a matrix product operator and optimizing the corresponding ground state 
wave function with imaginary-time time-dependent DMRG. 
We demonstrate our quantum chemical transcorrelated DMRG approach at the example
of several atoms and first-row diatomic molecules.
We show that transcorrelation improves the convergence rate to the complete basis set limit in comparison to conventional DMRG.
Moreover, we study extensions of our approach that aim at reducing the cost of handling the matrix product operator representation of the transcorrelated Hamiltonian.
\end{abstract}

\maketitle

\section{Introduction}
\label{sec:intro}

The reliable description of electron correlation in multi-configurational cases is one of the greatest challenges in modern quantum chemistry.
The accurate determination of the correlation energy is crucial for the reliability of theoretical predictions for many-electron atoms, molecules, and materials.

Electron correlation is usually divided into two contributions that have a distinct physical origin: static and dynamical correlation.
Static correlation is present in systems with a near-degenerate energy spectrum.
If energy levels are not well separated, a substantial number of configuration state functions with large weights are needed for a qualitatively reliable approximation of the wave function, signalling a breakdown of the mean-field picture.
By contrast, dynamical correlation embraces the numerous tiny contributions of configuration state functions constructed from weakly entangled orbitals.
In a first-quantized picture, the electron-electron cusp of the wave function in the vicinity of the coalescence of any two electrons\cite{kato57} is responsible for (short-range) dynamic electron correlation.
The electron-electron cusp feature is entirely absent in the mean-field description of the electronic wave function.
It is also difficult to describe with expansions in terms of antisymmetrized products of one-electron basis functions (i.e., orbitals), leading to the slow convergence of the dynamical-correlation energy with respect to basis set size~\cite{hill85}.

Despite the fact that static and dynamical correlation always coexist in any given system and are not rigorously separable in a strict sense, some molecular systems can be assigned to idealized situations where one type of correlation dominates. 
Efficient and general electronic structure methods have been developed to deal with such idealized situations.
In systems where the static correlation is negligible, coupled-cluster (CC) theory~\cite{crawford07,bartlett07} is capable of reaching chemical accuracy with polynomial-scaling cost. 
In particular, the explicitly correlated (F12) incantation~\cite{tenno2004,hattig12,kong12,werner2018} of the so-called ``gold standard'' CCSD(T) theory~\cite{ragha89} delivers an excellent efficiency-to-cost ratio.
As an alternative to the explicitly correlated approach, various extrapolation techniques~\cite{martin96,helgaker97,halkier98,varandas00,schwenke05,varandas07,feller11,lesiuk19} and corrections based on density functional theory~\cite{loos19,giner20} were proposed to eliminate basis set incompleteness errors.

In the idealized case where dynamical correlation is negligible but static correlation is strong, multi-configurational methods are required and models based on the full configuration interaction (FCI) approach are the most accurate and well defined.
Fortunately, it is very rare in molecular physics and chemistry to encounter systems where more than, say, 100 spatial orbitals are strongly correlated.
Efficient solutions of the FCI problem in this regime are offered by the density matrix renormalization group (DMRG),\cite{white92,white93,Chan2008_Review,Zgid2009_Review,Marti2010_Review-DMRG,Schollwoeck2011_Review,chan2011density,Wouters2013_Review,Kurashige2014_Review,Olivares2015_DMRGInPractice,szalay2015tensor,Yanai2015,Baiardi2020_Review} FCI quantum Monte Carlo~(FCIQMC)\cite{Alavi2009_FCIQMC-Original,cleland10}, and different flavors of selected CI.\cite{Holmes2016_SelectedCI,Evangelista2016_SelectedCI,Loos2018_CIPSI,HeadGordon2020_SelectedCI}

However, molecular systems in which both static and dynamical correlations are indispensable for accurate electronic energies and wave functions continue to pose a challenge, even for state-of-the-art electronic structure methods.
Conventional CC theory breaks down entirely, if a single, suitable reference determinant is not available. Methods have been devised to extend the range of applicability of CC theory without abandoning the single-reference framework such as externally corrected CC\cite{paldus84,paldus94,stolarczyk94,piecuch96,peris97,li97,aroeira2}, tailored CC\cite{kinoshita05,lyakh11,melnichuk12,melnichuk14,Veis2016_TCC,faulstich19,Morchen2020_TCC}, and method-of-moments approaches\cite{kowalski00a,kowalski00b,piecuch04,shen2012a,shen2012b,shen2012c,bauman2017}.
However, none of them represents a general solution to the quasidegeneracy problem (see also Ref.~\citenum{evangelista18}).

The reason for this situation is that 
various issues prevent the above mentioned modern strong-correlation methods from solving the dynamic correlation
problem to the same accuracy as the static correlation problem. For example,
dynamical correlation effects 
are difficult to capture with the one-dimensional matrix-product-state parameterization of DMRG
as DMRG resource requirements grow with the number
of spatial orbitals, restricting this number to about one hundred. Hence, most of the dynamical 
correlations are not accessible in a DMRG optimization of a matrix product state.
Such correlations also lead to FCI wave functions where many Slater determinants have a small, yet nonnegligible weight, which are hard to represent with FCIQMC.
A prominent approach to capture dynamical correlation effects is
multi-reference perturbation theory\cite{Yanai2013_DMRG-CASPT2,Roemelt2016_NEVPT2-DMRG,Freitag2017_NEVPT2,Booth2020_StochasticPT2}, which suffers from the fact that the final energy is no longer variational and only of low order, while
its calculation requires computationally expensive higher-order reduced density matrices.
It is, therefore, no surprise that the search for alternative approaches has been subject of 
intense research efforts and CC-based algorithms \cite{Veis2016_TCC,Morchen2020_TCC,Vitale2020_FCIQMC-TCC,Chan2021_ExternallyCorrected}
may be taken as a good example.
Strategies to include dynamical correlation effects by combining wave function-based calculations and density functional theory have also been designed.
The most prominent examples are the multiconfigurational pair\cite{LiManni2014-MCPDFT,Gagliardi2017_MCPDFT,Sharma2021_MCPDFT} and short-range\cite{Leininger1997_srDFT,Fromager2007_srDFT,Hedegard2015_srDFT-DMRG} density functional theories, which require a computational effort that is negligible compared to that of the CI-type calculation.
The availability of methods that can optimize full CI-type wave functions with up to 100 spatial orbitals ensures that static correlation effects can be included in wave function-based calculations for the vast majority of the molecules.
Still, the accuracy and universal applicability of methods that aim at including the missing dynamical correlation remain an open issue.

An alternative idea to tackle the dynamical correlation problem is to employ a transcorrelated~\cite{boiii69},
similarity-transformed Hamiltonian $e^{-F}\hat{H}\,e^F$.
As two operators related by a similarity transformation have the same spectrum (in the exact representation), the calculations based on the standard and a similarity-transformed Hamiltonian converge to the same answer.
By proper choice of the factor $F$, one can then address the Coulomb singularity of the interaction terms
in the Hamiltonian, which results in a cusp-less wave function that is easier to describe by conventional 
orbital expansions. In this way, short-range dynamical correlation can be efficiently described, alleviating
a significant part of the total dynamical correlation.

The idea of transcorrelated Hamiltonian was proposed by Boys and Handy decades ago~\cite{boiii69}, 
but the field remained dormant until a recent revival due to technical advances that address the emerging high-dimensional
integrals and the broken hermiticity of the Hamiltonian.
The fact that the similarity transformed Hamiltonian is non-hermitian 
does not allow for a straightforward variational minimization of the ground-state wave function.
In this respect, a key results was obtained by Luo and Alavi,\cite{Luo2018_tcFCIQMC-UEG} who showed that methods relying on the imaginary-time propagation can be applied to optimize the left and right eigenfunction of non-hermitian operators.
Accordingly, the FCIQMC method could be successfully combined with the transcorrelation approach.\cite{Luo2018_tcFCIQMC-UEG,Alavi2019_FermiHubbard-tcFCIQMC,Alavi2019_tcFCIQMC-Molecules,Guther2021_Be2-tcFCIQMC}
In a recent work,\cite{Baiardi2020_tcDMRG} we showed that the imaginary-time variant of the time-dependent DMRG method\cite{Verstraete2016_TangentSpace,Paeckel2019_TDDMRG-Review} can be effectively applied to the transcorrelated Hamiltonian as well.
The resulting transcorrelated DMRG (tcDMRG) significantly enhances DMRG convergence as demonstrated for the Fermi-Hubbard Hamiltonian.\cite{Baiardi2020_tcDMRG}

In this work, we extend the tcDMRG approach to solve the non-relativistic electronic Schr\"{o}dinger equation. Inspired by F12 electronic structure methods pioneered by Ten-no and collaborators\cite{tenno2004}, we propose to use a relatively simple factor $F$ which is dependent on the interelectronic distances only via a universal scalar function, $f(r_{ij})$. This is a different approach than in the recent papers of Alavi and collaborators\cite{Alavi2019_tcFCIQMC-Molecules,Guther2021_Be2-tcFCIQMC}, where a more elaborate and accurate Jastrow factor based on Schmidt-Moskowitz correlators\cite{schmidt90} was chosen. We opt here for a Ten-no-like correlator because of its simplicity, its transferability between atomic and molecular systems, and the rigorous elimination of the electron-electron cusp.
Moreover, we show that with the proposed correlator, the two- and three-body integrals entering the molecular transcorrelated Hamiltonian can be calculated analytically within the density fitting approximation.
We also extend our real-time electronic-structure DMRG algorithm\cite{Baiardi2020_ElectronDynamics} to imaginary time propagation, which then allows for the optimization of the right eigenfunction of the transcorrelated Hamiltonian.
We demonstrate that tcDMRG is more efficient than DMRG in two main aspects: first, it enhances the convergence of DMRG to the complete basis set limit, as expected from an explicitly correlated method.
Second, it reduces the correlation extent of the electronic wave function, which results in a more compact matrix product state.
This effect has been already observed by Alavi and co-workers~\cite{Alavi2019_FermiHubbard-tcFCIQMC,Guther2021_Be2-tcFCIQMC} in the context of the transcorrelated FCIQMC method, where the lower correlation degree results is a more compact full CI wave function.
In the following section, we present the theory of molecular tcDMRG and then turn to a discussion of numerical results for the helium and beryllium atoms and for the diatomics of hydrogen, nitrogen, and beryllium.

\section{Molecular Transcorreled Density Matrix Renormalization Group Theory}
\label{sec:theory}

\noindent In this section, we briefly review the transcorrelated Hamiltonian for the non-relativistic molecular electronic Schr\"{o}dinger equation.
Then, we describe how we optimize the right eigenvectors of the transcorrelated Hamiltonian with the imaginary-time variant of time-dependent DMRG theory, called iTD-DMRG.\cite{Baiardi2019_TD-DMRG}
We have already applied this strategy in Ref.~\citenum{Baiardi2020_tcDMRG} to the Fermi-Hubbard Hamiltonian.\cite{Ehlers2015_MomentumSpaceFermiHubbard}
Finally, we introduce the density fitting-based algorithm for the calculation
of the integrals that enter the second-quantized form of the transcorrelated Hamiltonian.

\subsection{The molecular transcorrelated Hamiltonian}
\label{sec:transcorrelatedHamiltonian}

Consider the standard non-relativistic electronic Hamiltonian for an arbitrary atomic or molecular system (in Hartree atomic units)

\begin{equation}
  \hat{H} = -\half \sum_i \nabla^2_i - \sum_{iA} \frac{Z_A}{r_{iA}} + \sum_{j<i}\frac{1}{r_{ij}} \, ,
  \label{eq:h0}
\end{equation}
where the indices $i$, $j$ denote electrons and $A$ denotes nuclei (for the sake of simplicity, we have
omitted the nucleus-nucleus repulsion contribution as it is easy to evaluate in Born-Oppenheimer approximation).
The transcorrelated method of Boys and Handy~\cite{boiii69} applies a non-hermitian similarity transformation to $\hat{H}$,

\begin{equation}
  \widetilde{H} = e^{-F}\hat{H}\,e^F \, ,
  \label{eq:SimilarityTransformedHamiltonian}
\end{equation}
where $F$ is a function symmetric with respect to exchange of all electrons and usually written as sum of pairwise terms:
\begin{equation}
  F = \sum_{j<i}f(r_{ij}) \, .
  \label{eq:CorrelationFactor}
\end{equation}
In analogy with explicitly correlated electronic structure methods~\cite{hattig12,kong12}, we refer to the scalar functions $f(r_{ij})$ as correlation factors.
By applying the nested commutator expansion (also known as Baker-Campbell-Hausdorff formula), the Hamiltonian in Eq.~(\ref{eq:SimilarityTransformedHamiltonian}) can be exactly rewritten as a sum of up to three-body terms
\begin{equation}
  \widetilde{H} = \hat{H} - \sum_{j<i} L_{ij} - \sum_{k<j<i} K_{ijk},
  \label{eq:TranscorrelatedHamiltonian}
\end{equation}
where the two-body potential $L_{ij}$ is defined as
\begin{equation}
  L_{ij} = \half \big[ \nabla^2_i + \nabla^2_j, f(r_{ij}) \big] + \big[ f'(r_{ij}) \big]^2,
  \label{eq:lij}
\end{equation}

where the commutator of two operators $X$ and $Y$ is denoted by $[X,Y]$ and $f'(r_{ij})$ is the derivative of $f(r_{ij})$.
The three-body potential $K_{ijk}$ reads
\begin{align}
    K_{ijk} = \nabla_i f(r_{ij} ) \cdot \nabla_i f(r_{ik} )
            + \nabla_j f(r_{ij} ) \cdot \nabla_j f(r_{jk} )
            + \nabla_k f(r_{jk} ) \cdot \nabla_k f(r_{ik} ),
    \label{eq:kijk}
\end{align}
and is symmetric with respect to exchange of all particles.
The main idea behind the transcorrelated method relies on two observations.
First, $\hat{H}$ and $\widetilde{H}$ are related by a similarity transformation and, therefore, they possess the same spectrum if diagonalized in the full Hilbert space.
Therefore, one can calculate exact full-CI energies by solving the eigenvalue problem for $\widetilde{H}$ 
rather than for $\hat{H}$.
Second, the correlation factor can be purposefully chosen to eliminate the Coulomb singularity, $1/r_{ij}$, from the transformed Hamiltonian $\widetilde{H}$.
This makes the right eigenfunctions of $\widetilde{H}$ free of the cusp at the electron-electron coalescence points~\cite{kato57,pack66} and, therefore, easier to represent with one-electron basis sets.
The Coulomb singularity is eliminated when the correlation factor behaves as 

\begin{equation}
  f(r_{12}) =\half r_{12} + \mathcal{O}(r_{12}^2)
  \label{eq:LimitCorrelationFactor}
\end{equation}
at small interelectronic distances ($r_{12} \rightarrow 0$).
This still leaves considerable freedom in the choice of $f(r_{12})$.
For this work, we chose 

\begin{equation}
  f(r_{12})=\half r_{12}\,e^{-\gamma r_{12}},
  \label{eq:CorrelationFactorForm}
\end{equation}
where $\gamma$ is an adjustable parameter.
This choice guarantees that the correlation factor decays exponentially at large interelectronic distances, therefore reducing the number of numerically significant matrix elements of $L_{ij}$ and $K_{ijk}$ for large systems. 

For the sake of clarity, we introduce an extended notation for two- and three-body electron-repulsion integrals
\begin{equation}
  (A|\hat{X}(r_{12})|B) = \iint d\rone\,d\rtwo\,A(\rone)\,\hat{X}(r_{12})\,B(\rtwo),
  \label{eq:CoulombNotationTwoBody}
\end{equation}
\begin{equation}
  (A|\hat{X}(r_{12})|B|\hat{Y}(r_{13})|C) = \iiint d\rone\,d\rtwo\,d\rthr\,
  A(\rone)\,\hat{X}(r_{12})\,B(\rtwo)\,\hat{Y}(r_{13})\,C(\rthr),
  \label{eq:CoulombNotationThreeBody}
\end{equation}
for all one-electron functions $A(\bfr)$, $B(\bfr)$, $C(\bfr)$, and operators $\hat{X}$, $\hat{Y}$.
We denote the elements of an atomic-orbital (AO) basis by Greek lowercase letters, $\mu$, $\nu$, $\lambda$, etc.
Moreover, we take all such one-electron basis functions as real-valued functions. 

By inspection of Eqs.~(\ref{eq:lij}) and (\ref{eq:kijk}) one can show that the following integrals are needed in 
addition to the standard integrals for calculations with the transcorrelated Hamiltonian

\begin{align}
    \label{eq:intt}
    L_{\mu\nu,\lambda\sigma}^{(1)}      &= (\mu\nu| \Big[ \nabla_1^2, f(r_{12}) \Big]|\lambda\sigma), \\
    \label{eq:intu}
    L_{\mu\nu,\lambda\sigma}^{(2)}      &= (\mu\nu| \big[f'(r_{12})\big]^2 |\lambda\sigma), \\
    \label{eq:intw}
    K_{\mu\nu,\lambda\sigma,\kappa\tau} &= (\mu\nu|\nabla_1 f(r_{12} )|\lambda\sigma|\nabla_1 f(r_{13})|\kappa\tau).
\end{align}
The remaining terms in Eqs.~(\ref{eq:lij}) and (\ref{eq:kijk}) are obtained by permutation of orbital indices or exchange of electronic coordinates.
Whereas the additional two-electron integrals of Eqs.\ (\ref{eq:intt}) and (\ref{eq:intu}) do not compromise computational
feasibility,
the main hurdle is the three-body part in Eq.~(\ref{eq:intw}).
Assuming that the AO basis is of dimension $N$, the cost of calculating $(\mu\nu|\nabla_1 f(r_{12} )|\lambda\sigma|\nabla_1 f(r_{13})|\kappa\tau)$ is asymptotically proportional to $N^6$.
Even worse, quantum chemical calculations usually rely on a molecular or a local basis, rather than the atomic basis.
Transformation of the three-body integrals from one basis to another incurs $N^7$ computational costs with a large prefactor. 

In this work, we circumvent the bottleneck of the three-body-integral evaluation by density fitting, i.e., by resolution of the identity approximations.
The six-index integrals, Eq.~(\ref{eq:intw}), are decomposed into combinations of three-index quantities that can be evaluated analytically.
Transformations from one basis to another then become trivial and the integrals can be reconstructed directly in the molecular orbital basis if needed.
This is advantageous from the perspective of methods such as DMRG, where the size of the molecular orbital basis, \textit{i.e.} the active space, is typically much smaller than $N$. 
Note that our analytical approach differs from the one adopted by Cohen et al.\cite{Alavi2019_tcFCIQMC-Molecules} in two major ways.
First, we apply a relatively simple function $F$ in Eq.~(\ref{eq:CorrelationFactor}) that is responsible solely for elimination of the electron-electron cusp and aimed at capturing the dynamic correlation effects.
By contrast, in the work of Cohen et al., a more extensive correlation factor is used that additionally includes electron-nucleus and three-electron terms. 
While such a factor can be expected to provide more accurate energies, its main drawback is the emergence of complicated non-standard two- and three-body multicenter integrals which must be resolved with a six-dimensional numerical quadrature.
This is avoided by our choice for $F$.

\subsection{Transcorrelated Density Matrix Renormalization Group}
\label{sec:tcDMRG}

\noindent The two-body part of the transcorrelated Hamiltonian is non-hermitian.
Hence the corresponding eigenvalue problem cannot be solved with standard methods based on the variational principle.
This problem has been addressed with two classes of approaches.
The first class~\cite{boiii69} adopts biorthogonal variational principle, where the basis sets in which the right and left eigenvectors of the transcorrelated Hamiltonian are expanded are distinct.
Recently, a similar idea was exploited by Ten-no~\cite{hino02,Hino2001_Biorthogonal-tcCC} in the context of transcorrelated coupled-cluster theory.
The second approach to deal with the lack of hermiticity of the transcorrelated Hamiltonian is based on imaginary-time evolution starting with a reasonable approximation for the exact ground state. 
As shown by Luo and Alavi~\cite{Luo2018_tcFCIQMC-UEG}, this approach does not suffer from instabilities associated with the lack of hermiticity of the Hamiltonian.
Inspired by these findings, we have introduced~\cite{Baiardi2020_tcDMRG} transcorrelated DMRG (tcDMRG) that optimizes the ground-state of the transcorrelated Hamiltonian with iTD-DMRG.
The approach adopted in the present work is an extension of this formalism.

\noindent The tcDMRG algorithm\cite{Baiardi2020_tcDMRG} encodes the transcorrelated Hamiltonian as a matrix product operator (MPO) and its ground-state eigenfunction as a matrix product state (MPS).\cite{McCulloch2007_MPS}
The MPS representation of a FCI wave function in a basis of $L$ spatial orbitals 
reads:

\begin{equation}
  \ket{\Psi_\text{FCI}} = \sum_{\bm{\sigma}} C_{\bm{\sigma}} \ket{\bm{\sigma}}
        = \sum_{\bm{\sigma}} \sum_{a_1,\ldots,a_{L-1}}^m M_{1,a_1}^{\sigma_1} M_{a_1,a_2}^{\sigma_2}
                             \cdots M_{a_{L-1},1}^{\sigma_L} \ket{\bm{\sigma}} \, .
  \label{eq:MPS}
\end{equation}

\noindent Eq.~(\ref{eq:MPS}) expresses the full-CI tensor $C_{\bm{\sigma}}$ as the product of $L$ three-dimensional tensors of size $4 \times m \times m$, where a fixed value of $m$, known as the bond dimension, introduces an approximation that
tames the curse of dimensionality.
Accordingly, the matrix operator form of the Hamiltonian $\mathcal{H}$ also features such a local,
orbital-focused decomposition and reads in its exact form:

\begin{equation}
  \mathcal{H} = \sum_{\bm{\sigma},\bm{\sigma'}} \sum_{b_1,\ldots, b_{L-1}}^{b_\text{max}}
                    H_{1,b_1}^{\sigma_1, \sigma_1'} H_{b_1,b_2}^{\sigma_2, \sigma_2'}
                    \cdots H_{b_{L-1},1}^{\sigma_L,\sigma_L'} \ket{\bm{\sigma}} \bra{\bm{\sigma}'} \, .
  \label{eq:MPO}
\end{equation}

\noindent In this work, we encode the transcorrelated Hamiltonian as in Eq.~(\ref{eq:MPO}) starting from its second-quantized form, which reads:

\begin{equation}
 \begin{aligned}
  \mathcal{H}_\text{TC} =& \sum_{\mu \nu} \sum_{s}^{\left\{ \alpha,\beta \right\}}
        h_{\mu \nu} a_{\mu s}^+ a_{\nu s}
      + \frac{1}{2} \sum_{\mu \nu \lambda \sigma} \sum_{s s'}^{\left\{ \alpha,\beta \right\}}
        \left[ ( \mu \nu | \lambda \sigma ) - L_{\mu\nu,\lambda\sigma}^{(1)} - L_{\mu\nu,\lambda\sigma}^{(2)} \right]
        a_{\mu s}^+ a_{\lambda s'}^+ a_{\sigma s'} a_{\nu s} \\
    &- \frac{1}{6} \sum_{\mu \nu \lambda \sigma \kappa \tau} \sum_{s s' s''}^{\left\{ \alpha,\beta \right\}}
       K_{\mu\nu,\lambda\sigma,\kappa\tau} 
       a_{\mu s}^+ a_{\lambda s'}^+ a_{\kappa s''}^+ a_{\tau s''} a_{\sigma s'} a_{\nu s}
 \end{aligned}
 \label{eq:TC_SecondQuantization}
\end{equation}

\noindent We encode Eq.~(\ref{eq:TC_SecondQuantization}) by applying the algorithm introduced in Ref.~\citenum{Keller2015_MPS-MPO-QuantumChemical} that we generalized in Ref.~\citenum{Baiardi2020_tcDMRG} to operators that include three-body terms.
We note that, formally, applying DMRG to Hamiltonians with long-range interactions and multi-index parameters, 
such as the transcorrelated one, is straightforward within the MPO/MPS formulation of the DMRG~\cite{Keller2015_MPS-MPO-QuantumChemical,Chan2016_DifferentDMRGFormulations,Baiardi2020_DMRG-Perspective}.
In fact, the algorithm introduced in Ref.~\citenum{Keller2015_MPS-MPO-QuantumChemical} can support strings of second-quantized operators of arbitrary size.
Instead, encoding the three-body term would be extremely challenging within the original formulation of DMRG due to the need to calculate explicitly the matrix element of each possible combination of elementary second quantization operators over the left- and right-renormalized basis.
We also note that the bond dimension $b_i$ (see Eq.~(\ref{eq:MPO})) of the MPO representation of Eq.~(\ref{eq:TranscorrelatedHamiltonian}) scales as $\mathcal{O}(L^5)$~\cite{Crosswhite2008_FiniteAutomata}.
This scaling determines the storage requirements for the boundaries that collect the result of the partial MPS/MPO contractions,\cite{Schollwoeck2011_Review} which is the main bottleneck of our current implementation of the tcDMRG method. 

As has already been mentioned above, the two-body part of the transcorrelated Hamiltonian is non-hermitian.
Therefore, the right eigenfunctions cannot be optimized with conventional DMRG.
Van Voorhis and Chan generalized the time-independent DMRG (TI-DMRG) method to non-hermitian operators~\cite{VanVoorhis2005_NonHermitian-DMRG,Helms2019_LargeDeviations-DMRG}.
This approach relies on the bivariational principle and encodes both the left and right Hamiltonian eigenfunctions as MPSs.
However, the key advantage of transcorrelated approaches is that they reduce the correlation degree of the right eigenfunction, which can therefore be encoded as a compact MPS, at the price of increasing the complexity of the left one.\cite{Alavi2019_FermiHubbard-tcFCIQMC}
Approaches that represent only the right eigenfunction, but not the left one, will benefit from the transcorrelation approach.
This is the case for methods based on imaginary-time propagation, such as FCIQMC\cite{Alavi2009_FCIQMC-Original,Luo2018_tcFCIQMC-UEG,Alavi2019_FermiHubbard-tcFCIQMC} and iTD-DMRG.\cite{Corboz2011_iPEPS,Orus2012_iTD-CornerTensors,Vidal2015_ImaginaryTime-TNS,Baiardi2020_ElectronDynamics,Baiardi2020_tcDMRG}
Among the TD-DMRG variants that have been proposed in recent years,\cite{Paeckel2019_TDDMRG-Review} we rely on the tangent-space TD-DMRG method\cite{Lubich2015_TimeIntegrationTT,Verstraete2016_TangentSpace} that simulates the imaginary-time evolution of an MPS by projecting the corresponding time-dependent Schr\"{o}dinger equation onto the manifold of the MPSs with a given bond dimension $m$, \textit{i.e.}:

\begin{equation}
  \frac{\text{d}\ket{\Psi_\text{MPS}(t)}}{\text{d}t} 
    = - \mathcal{P}_{\Psi_\text{MPS}} \mathcal{H}_\text{TC} \ket{\Psi_\text{MPS}(t)} \, ,
  \label{eq:TangentSpace}
\end{equation}
$\mathcal{P}_{\Psi_\text{MPS}}$ being the so-called tangent-space projector which ensures that $\mathcal{H}_\text{TC} \ket{\Psi_\text{MPS}}$ is expressed as an MPS with bond dimension $m$.
Naturally, this projection operation approximates the exact imaginary-time evolution.
For real-time propagation, this approximation will have a negligible effect on the accuracy 
of the calculation if the wave function can be represented as an MPS of bond dimension $m$ at all times.
It is sufficient for this condition not to be met at a single instant in time 
in order to compromise the accuracy of the whole subsequent propagation.
Conversely, this approximation is not critical for imaginary-time propagations. 
In fact, the imaginary-time evolution yields the optimal MPS only in the infinite-propagation-time limit $t \rightarrow +\infty$.
It is, therefore, sufficient that the energy is converged with respect to
$m$ for $t \rightarrow +\infty$ to ensure the tcDMRG convergence. 

We solve Eq.~(\ref{eq:TangentSpace}) with a second-order Trotter factorization of the imaginary-time propagator\cite{Lubich2015_TimeIntegrationTT,Verstraete2016_TangentSpace} as described in Ref.~\citenum{Baiardi2020_tcDMRG}.
In practice, the $M_{a_{i-1},a_i}^{\sigma_i}$ tensors are propagated sequentially, one after the other, with a sweep-based scheme that is equivalent to the one of conventional TI-DMRG.
Note that the Trotter approximation is the second error source that is introduced in addition to the truncation error.
For this reason, the tcDMRG energy convergence must be monitored with respect to both the bond dimension $m$ and the propagation time-step $\Delta t$. \\
In the following, we denote the algorithm described above as single-site tcDMRG (tcDMRG[SS]).
The algorithm can be modified to propagate, at each microiteration, the two-site tensor $T_{a_{i-1},a_{i+1}}^{\sigma_i,\sigma_{i+1}}$ obtained by contracting the MPS tensors centering on two neighbouring sites $(i)$ and ($i$+1).
The resulting tcDMRG variant, namely tcDMRG[TS],\cite{Verstraete2016_TangentSpace} is less prone to converge to local minima of the energy functional.
Moreover, it enables dynamically adapting the bond dimension $m$ to the wave function entanglement, as is done in the dynamical block state selection method.\cite{Legeza2003_DBSS}
As we will show in Section~\ref{sec:results}, tcDMRG[TS] is the most accurate optimization algorithm for strongly correlated molecules.

We conclude this section by highlighting the formal scaling of tcDMRG with respect to its key parameters, \textit{i.e.} the number of orbitals $L$, the bond dimension $m$, and the number of sweeps $N_\text{sweeps}$ for the imaginary-time evolution.
We first note that the imaginary-time evolution of the MPS wave function with iTD-DMRG is the main bottleneck of tcDMRG because, owing to the density fitting approximation that we will describe in the next section, the calculation of the three-body integrals is characterized by a negligible computational cost.
The scaling of the imaginary-time evolution is $\mathcal{O}(N_\text{sweeps} L N_\text{site})$, $N_\text{site}$ being the computational scaling of a microiteration step of iTD-DMRG.
We approximate the evolution of a single MPS tensor with $N_\text{Lanczos}$ iterations of the Lanczos algorithm.
Therefore, $N_\text{site} = N_\text{Lanczos} N_\text{contr}$, where $N_\text{contr}$ is the computational cost associated with the contraction of the MPS tensor with the MPO tensor and the boundary.
The bond dimension of the MPO representation of the transcorrelated Hamiltonian scales as $\mathcal{O}(L^5)$.
Therefore, following Ref.~\citenum{Keller2015_MPS-MPO-QuantumChemical}, the scaling of the contraction is $\mathcal{O}(m^3L^5) + \mathcal{O}(m^2L^6)$, with a prefactor that depends on the sparsity of the three-electron integral tensor.

\subsection{Evaluation of the matrix elements}
\label{sec:denfit}

Besides the lack of hermiticity problem addressed in the previous section, the second major challenge of the transcorrelated methods is the need to calculate non-standard two- and three-body integrals.
Moreover, as these integrals are usually calculated in the atomic orbital basis, they must additionally be transformed to the molecular orbital basis to facilitate DMRG calculations.
Both these steps are computationally demanding since the three-body integrals are six-index quantities.
In the present work we leverage the density fitting approximation to tame the high computational cost of these two steps.

In the density-fitting approximation~\cite{whitten73,baerends73,dunlap79,alsenoy88,vahtras93} a product of two basis functions is expanded into a linear combination of pre-optimized auxiliary basis functions.
The elements of the latter basis shall be denoted by capital letters $P$, $Q$, etc.
This leads to
\begin{equation}
  |\mu\nu) \approx \sum_P C_{\mu\nu}^P\,|P) \, .
  \label{eq:df1}
\end{equation}
The coefficients $C_{\mu\nu}^P$ are obtained by minimizing the error of Eq.~(\ref{eq:df1}) in the square norm under a certain metric. In this work we employ the Coulomb metric where the error $\epsilon_{\mu\nu}$ takes the form
\begin{equation}
\epsilon_{\mu\nu} = 
  \iint d\rone\,d\rtwo\, r_{12}^{-1} \Big[ |\mu\nu) - \sum_P C_{\mu\nu}^P\,|P) \Big]^2.
  \label{eq:df2}  
\end{equation}
Minimization with respect to the coefficients leads to the following set of linear equations
\begin{equation}
  \sum_{Q} (P|r_{12}^{-1}|Q)\,C_{\mu\nu}^Q = (\mu\nu|r_{12}^{-1}|P) \, ,
  \label{eq:dfcoefs1}
\end{equation}
which can easily be solved. Introducing a compact notation for the basic integrals
\begin{equation}
  J_{\mu\nu}^P = (\mu\nu|r_{12}^{-1}|P) 
  \label{eq:dfBasicIntegrals1}
\end{equation}
and
\begin{equation}
  V_{PQ} = (P|r_{12}^{-1}|Q) \, ,
  \label{eq:dfBasicIntegrals2}
\end{equation}
we can explicitly write the solution of Eq.~(\ref{eq:dfcoefs1}) as
\begin{equation}
  C_{\mu\nu}^Q = \sum_P J_{\mu\nu}^P \big[\mathbf{V}^{-1}\big]_{PQ}.
  \label{eq:dfSolution}
\end{equation}
Under the approximation introduced in Eq.~(\ref{eq:df1}), the standard Coulomb integrals read
\begin{equation}
  (\mu\nu|r_{12}^{-1}|\lambda\sigma) \approx C_{\mu\nu}^P\,J_{\lambda\sigma}^P \, .
  \label{eq:dfCoulombIntegrals}
\end{equation}

One can show that this formula is automatically ``robust'' in the sense that the error in the integrals is quadratic in the density error~\cite{dunlap00} defined by Eq.~(\ref{eq:df1}).
This is the case because the fitting metric is the same as the two-electron kernel in the integrals ($r_{12}^{-1}$ in our case).
A more general formula that is ``robust'' for an arbitrary multiplicative kernel $\hat{X}(r_{12})$ reads
\begin{equation}
  (\mu\nu|\hat{X}(r_{12})|\lambda\sigma) \approx
    C_{\mu\nu}^P\,X_{\lambda\sigma}^P +
    C_{\lambda\sigma}^P\,X_{\mu\nu}^P -
    C_{\mu\nu}^P\,X_{PQ}\,C_{\lambda\sigma}^Q,
  \label{eq:dfrobust}
\end{equation}
where
\begin{equation}
  X_{\mu\nu}^P = (\mu\nu|\hat{X}(r_{12})|P),\;\;\;
    X_{PQ} = (P|\hat{X}(r_{12})|Q).
  \label{eq:dfRobust2}
\end{equation}
It has been shown that the ``robust'' formulation is more accurate than other variants of the density fitting approximation~\cite{vahtras93}. \\
Eq.~(\ref{eq:dfrobust}) can be straightforwardly applied to the basic integrals in Eq.~(\ref{eq:intu}) by simply setting $\hat{X}(r_{12})=\big[f'(r_{12})\big]^2$.
The two-index and three-index integrals needed in this case are
\begin{equation}
    G_{\mu\nu}^P = (\mu\nu|\big[f'(r_{12})\big]^2|P),\;\;\;
    G_{PQ} = (P|\big[f'(r_{12})\big]^2|Q).
\end{equation}
and can be evaluated analytically.
The second class of two-electron integrals, Eq.~(\ref{eq:intt}), is more problematic from the density-fitting perspective because the kernel $\Big[ \nabla_1^2, f(r_{12}) \Big]$ is not multiplicative.
The simplest solution is to approximate the integrals of Eq.~(\ref{eq:intt}) with a non-robust formula, although this may degrade the accuracy significantly.
We circumvent this problem by following Manby~\cite{manby03} and introduce an antisymmetric density-like product
\begin{equation}
  |\{\mu\nu\}) = |\big(\nabla^2 \mu\big) \nu - \mu\big(\nabla^2 \nu\big)).
\end{equation}
By applying the divergence theorem, Eq.~(\ref{eq:intt}) can be rewritten into a fully equivalent form
\begin{equation}
  (\mu\nu| \Big[ \nabla_1^2, f(r_{12}) \Big]|\lambda\sigma) =
    (\{\mu\nu\}| f(r_{12}) |\lambda\sigma) \, .
\end{equation}
The advantage of this formulation is that the quantity $|\{\mu\nu\})$ is a one-electron object that can be approximated with the density-fitting approach.
To this end, we write
\begin{equation}
  |\{\mu\nu\}) \approx \sum_P \bar{C}_{\mu\nu}^P\,|P) \, ,
\end{equation}
and the coefficients are again found by minimizing the error in the Coulomb metric:
\begin{equation}
  \overline{C}_{\mu\nu}^Q = \sum_P \bar{J}_{\mu\nu}^P \big[\mathbf{V}^{-1}\big]_{PQ} \, ,
\end{equation}
where
\begin{equation}
  \bar{J}_{\mu\nu}^P = (\{\mu\nu\}|r_{12}^{-1}|P) \, .
\end{equation}
Fitting $|\{\mu\nu\})$ may be more difficult than fitting $|\mu\nu)$ and may require a larger auxiliary basis.
This issue shall be investigated numerically in the next section.
It is now possible to write the ``robust'' density-fitted formula for the remaining integrals
\begin{equation}
  (\{\mu\nu\}| f(r_{12}) |\lambda\sigma) \approx
    \overline{C}_{\mu\nu}^P\,F_{\lambda\sigma}^P +
    \overline{F}_{\mu\nu}^P\,C_{\lambda\sigma}^P -
    \overline{C}_{\mu\nu}^P\,F_{PQ}\,C_{\lambda\sigma}^Q \, ,
\end{equation}
where
\begin{equation}
  F_{\mu\nu}^P = (\mu\nu|f(r_{12})|P) \, , \;\;\;
    \overline{F}_{\mu\nu}^P = (\{\mu\nu\}|f(r_{12})|P) \, , \;\;\;
    F_{PQ} = (P|f(r_{12})|Q) \, .
\end{equation}
Passing to the three-body component, we first point out that a ``robust'' density-fitting formula for three-electron integrals is also available~\cite{womack14}.
In the case of the basic integrals of Eq.~(\ref{eq:intw}) it reads
\begin{equation}
 \begin{aligned}
    (\mu\nu|&\nabla_1 f(r_{12} )|\lambda\sigma|\nabla_1 f(r_{13})|\kappa\tau) \approx \\
    &C_{\mu\nu}^P\,C_{\lambda\sigma}^Q\,\alpha_{PQ}^{\kappa\tau} +
    C_{\mu\nu}^P\,C_{\kappa\tau}^Q\,\alpha_{PQ}^{\lambda\sigma} +
    C_{\lambda\sigma}^P\,C_{\kappa\tau}^Q\,\beta_{PQ}^{\mu\nu} -
    2 C_{\mu\nu}^P\,C_{\lambda\sigma}^Q\,C_{\kappa\tau}^R\,\gamma_{PQR},
 \end{aligned}
 \label{eq:df3}
\end{equation}
with
\begin{align}
  \alpha_{PQ}^{\kappa\tau} &= (P|\nabla_1 f(r_{12} )|Q|\nabla_1 f(r_{13})|\kappa\tau), 
  \label{eq:alpha0} \\
  \beta_{PQ}^{\kappa\tau}  &= (\mu\nu|\nabla_1 f(r_{12} )|P|\nabla_1 f(r_{13})|Q), 
  \label{eq:beta0} \\
  \gamma_{PQR} &= (P|\nabla_1 f(r_{12} )|Q|\nabla_1 f(r_{13})|R).
  \label{eq:gamma0}
\end{align}
In analogy to Eq.~(\ref{eq:dfrobust}), this formula guarantees that the error in the integrals is quadratic in the fitting error in Eq.~(\ref{eq:df1}).
Eq.~(\ref{eq:df3}) conveniently reduces the initial six-index integral to a combination of only three- and four-index quantities.
The new basic three-electron integrals $\alpha_{PQ}^{\kappa\tau}$, $\beta_{PQ}^{\kappa\tau}$, and $\gamma_{PQR}$ can be evaluated analytically as in the paper of Womack and Manby~\cite{womack14} or Ten-no~\cite{tenno00}.
The computational cost would, however, remain high because some four-index integrals involve two elements of the density-fitting basis, which is usually several times larger than the corresponding AO basis.
On top of that, these integrals must be transformed to the molecular or local orbital basis.

We therefore propose to eliminate the three-electron integrals by proper insertion of the resolution of the identity (RI)
\begin{equation}
  \label{eq:rimain}
  \hat{1} \approx |x\rangle\langle x|,
\end{equation}
where $|x\rangle$ are elements of an orthonormal RI basis.
We remove all four-index quantities from the integrals of Eqs.~(\ref{eq:alpha0})--(\ref{eq:gamma0}) by applying the RI (Eq.~(\ref{eq:rimain})).
The final formulae read
\begin{align}
    \alpha_{PQ}^{\kappa\tau} &= -\big(\nabla[Px]\big|f(r_{12})\big|Q\big)\, 
    \big(\nabla[\kappa\tau]\big|f(r_{12})\big|x\big), \\
    \beta_{PQ}^{\kappa\tau}  &= +\big(\nabla[\kappa x]\big|f(r_{12})\big|Q\big)\, 
    \big(\nabla[\tau x]\big|f(r_{12})\big|P\big), \\
    \gamma_{PQR} &= -\big(\nabla[Px]\big|f(r_{12})\big|Q\big)\, 
    \big(\nabla R\big|f(r_{12})\big|x\big) \, ,
\end{align}
where we leveraged the divergence theorem to remove the derivatives of the correlation factor and, instead, differentiate the basis functions. 
The derivation of the expressions above is nontrivial and involves several steps which are explained in detail in 
the appendix.

We evaluate the new two-index, $\big(\nabla R\big|f(r_{12})\big|x\big)$, and three-index, $\big(\nabla[Px]\big|f(r_{12})\big|Q\big)$, integrals analytically with the McMurchie-Davidson scheme~\cite{murchie78,helgaker92}.
While this method may not be as efficient as more advanced and heavily optimized techniques, it is very general and offers a high degree of flexibility.
The correlation factor $f(r_{12})$ and its derivative $f'(r_{12})$ are separately expanded into a linear combination of pre-optimized Gaussian functions.
This allows for encoding virtually any short-range (local) correlation factor within the same implementation.
It is necessary to supply only Gaussian exponents and contraction coefficients of the fit.

As a final note we justify the use of the RI insertion for the three-body integrals, Eqs.~(\ref{eq:alpha0})--(\ref{eq:gamma0}).
It was shown by Kutzelnigg and Klopper~\cite{kutz91} that for this type of integrals the RI expansion truncates after a certain angular momentum for one-center systems.
In our particular case, it is sufficient to include in the RI basis angular momentum functions up to $2l_{\mathrm{max}}+1$, where $l_{\mathrm{max}}$ is the maximum angular momentum in the orbital basis.
Assuming that the basis is radially saturated, the RI expansion is exact.
Of course, this is no longer true for arbitrary molecules, but the truncation error introduced by evaluation of Eqs.~(\ref{eq:alpha0})--(\ref{eq:gamma0}) by RI should be rather small.

\section{Results}
\label{sec:results}

In this section, we apply the tcDMRG method described above to calculations for atomic and molecular systems.
First, based on results for helium and beryllium atoms, we study the accuracy of density fitting and resolution of the identity approximation of the two- and three-body transcorrelated integrals.
Next, we demonstrate for the hydrogen molecule that in comparison with conventional DMRG, tcDMRG results converge faster to the limit of a complete basis set.
Finally, we highlight strengths and limitations of the tcDMRG algorithm by applying it in calculations of the potential energy curves (PECs) of two molecular systems, namely Be$_2$ and N$_2$.

\subsection{Accuracy of the matrix elements}
To verify that the approach for evaluating the transcorrelated matrix elements presented in this work is correct, we calculate the energy of two few-body systems~--~the helium atom and the beryllium atom, with tcDMRG.
By comparing tcDMRG energy with near-exact values taken from the literature~\cite{nakashima07,puchalski13} for these atoms, we assess the effect of the adopted approximations on the accuracy of the results and highlight the key advantages of the tcDMRG method.
To this end, we calculated the absolute energy of He and Be with cc-pV$X$Z basis sets within the full orbital space with both TI-DMRG and tcDMRG.
All energies reported in this section are converged with the bond dimension $m$ and, therefore, both TI-DMRG and tcDMRG are factually equivalent to full CI.

\begin{table}[htbp!]
  \centering
  \caption{TI-DMRG and tcDMRG energies for the helium atom (Hartree atomic units) within the full orbital space.
  The reference FCI basis set limit is $-2.903\,724$ from Ref.~\citenum{nakashima07}.}
  \vspace{0.2cm}
  \begin{tabular}{lcc}
    \hline\hline
    Basis   & TI-DMRG & tcDMRG \\
    \hline
    cc-pVDZ & $-$2.889\,583 & $-$2.895\,728 \\
    cc-pVTZ & $-$2.900\,502 & $-$2.902\,531 \\
    cc-pVQZ & $-$2.902\,552 & $-$2.903\,379 \\
    cc-pV5Z & $-$2.903\,188 & $-$2.903\,629 \\
    \hline\hline
  \end{tabular}
  \label{tab:table_he_trans}
\end{table}

\begin{table}[htbp!]
  \centering
  \caption{TI-DMRG and tcDMRG energies for the beryllium atom (Hartree atomic units) within the full orbital space.
  The reference FCI basis set limit is $-14.667\,356$ from Ref.~\citenum{puchalski13}.}\vspace{0.2cm}
  \begin{tabular}{cccc}
    \hline\hline
    Basis   & TI-DMRG & tcDMRG         & tcDMRG \\
            &         & (without 3-body)  & (full Hamiltonian) \\
    \hline
    cc-pVDZ & $-$14.617\,572 & $-$14.650\,738 & $-$14.656\,855 \\
    cc-pVTZ & $-$14.623\,832 & $-$14.652\,750 & $-$14.658\,854 \\
    cc-pVQZ & $-$14.640\,169 & $-$14.657\,769 & --- \\
    \hline\hline
  \end{tabular}
  \label{tab:table_be_trans}
\end{table}

In the complete basis set limit both TI-DMRG and tcDMRG energies converge to the exact solution of the electronic Schr\"odinger equation.
The matching cc-pV$X$Z-RIFIT basis sets of Weigand \emph{et al.}~\cite{weigend98} were employed for the density-fitting approximation, while the uncontracted cc-pV6Z basis was used for RI.
The results are given in Tables~\ref{tab:table_he_trans} and \ref{tab:table_be_trans} for He and Be, respectively.
In the case of the beryllium atom with cc-pVQZ basis set tcDMRG energies are not available due to exceeding memory requirements.
For all calculations reported in the present work, if not explicitly stated otherwise,  we set the parameter~$\gamma$ in the correlation factor [$f(r_{12})=\half r_{12}\,e^{-\gamma r_{12}}$] to 1.

It is clear that the tcDMRG energies converge to the basis set limit significantly faster than the TI-DMRG ones.
For instance, the tcDMRG/cc-pVTZ energy of He is lower than the TI-DMRG/cc-pVQZ one.
These gains are even larger for the beryllium atom where the three-body terms enter the transcorrelated Hamiltonian, as opposed to the helium atom.
We report in Table~\ref{tab:table_be_trans} the Be tcDMRG energies obtained by including only the two-body contributions to the transcorrelated Hamiltonian, \textit{i.e.} by neglecting the pure three-body operator, Eq.~(\ref{eq:kijk}).
We refer to this approximate method as tcDMRG[2].
The motivation for considering such an approximation is the fact that the three-body part of the transcorrelated Hamiltonian is responsible for the aforementioned memory bottleneck and limits the size of the systems that can be treated at present with tcDMRG.
Unfortunately, the contribution of the three-body term is substantial (several mHa) and its neglect cannot be justified based solely on the energy criterion.
Moreover, there is no indication that the magnitude of this contribution decreases when the basis set is enlarged.

One may argue that the results given in Table~\ref{tab:table_be_trans} for the beryllium atom still include some residual error in comparison with the exact limit and with another variant of the transcorrelated approach put forward by Alavi and collaborators~\cite{Guther2021_Be2-tcFCIQMC}. However, the reason for this discrepancy of almost 10 mHa is the fact that the correlation factor of Eq.~(8) does not eliminate the electron-nuclear cusp from the Hamiltonian, in contrast with Ref.~\citenum{Guther2021_Be2-tcFCIQMC}. 
As the main focus of this work is not to deal with the nuclear-electron cusp, for which we believe that there are more efficient methods (other than transcorrelation) available, we do not further discuss this difference in total energy.
Instead, we focus on including the dynamical electron correlation into genuine multireference quantum chemistry methods, which is a problem that has no satisfactory solution at present.

\begin{table}[htbp!]
  \centering
  \caption{Stability of the tcDMRG energies for the beryllium atom (cc-pVDZ orbital basis set) with respect to the size of the density-fitting (DF) basis set (in rows) and RI basis set (in columns).} \vspace{0.2cm}
  \begin{tabular}{c|ccc}
    \hline\hline
               & \multicolumn{3}{c}{RI basis (uncontracted)} \\
    DF basis   & cc-pVQZ & cc-pV5Z & cc-pV6Z \\
    \hline
    cc-pVDZ-RIFIT & $-$14.656\,869 & $-$14.656\,855 & $-$14.656\,808 \\
    cc-pVTZ-RIFIT & $-$14.656\,724 & $-$14.656\,696 & $-$14.656\,643 \\
    cc-pVQZ-RIFIT & $-$14.656\,727 & $-$14.656\,703 & $-$14.656\,647 \\
    \hline\hline
  \end{tabular}
  \label{tab:table_be_stability}
\end{table}

Finally, we consider the effect on the density-fitting and RI approximations on the accuracy of the tcDMRG energies.
We report in Table~\ref{tab:table_be_stability} tcDMRG energies for the beryllium atom with a fixed atomic orbital basis set (cc-pVDZ) and with increasingly large density-fitting and RI basis sets.
The convergence pattern indicates that the standard density-fitting basis (cc-pVDZ-RIFIT) yields an accuracy of $0.1-0.3$~mHa.
We expect this to be sufficient for most purposes, especially taking into consideration that a significant portion of the density-fitting error cancels out in calculations of relative energies.
If the density-fitting basis set is enlarged by one cardinal number, the errors will be within the $\mu$Ha range.
While the convergence of the results with respect to the size of the RI basis is somewhat less regular, the differences are again within $\mu$Ha range.

\subsection{Basis set convergence of the H$_2$ molecule}
\label{subsec:H2}

\begin{figure}[htbp!]
  \centering
  \includegraphics[width=.9\textwidth]{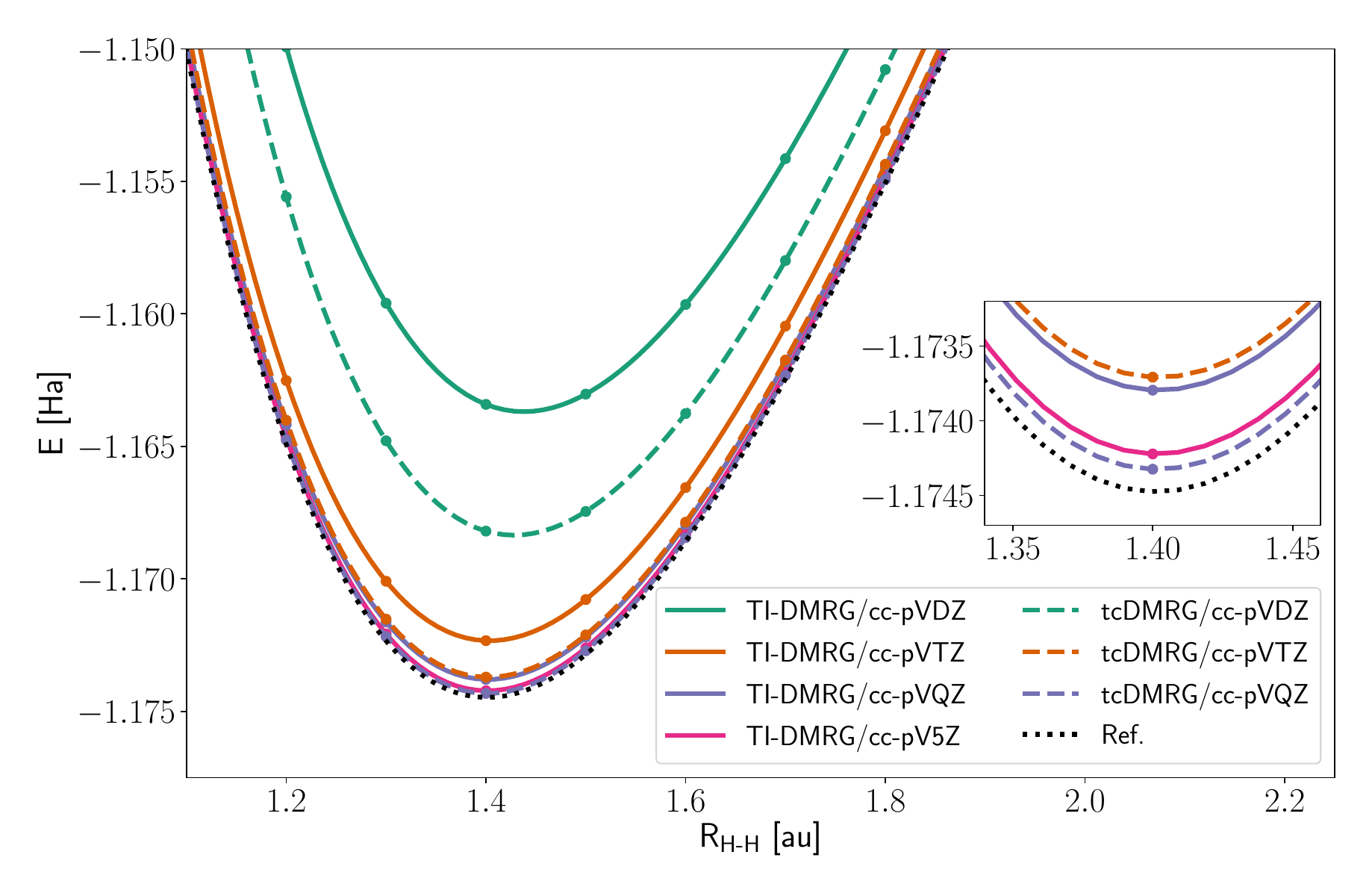}
  \caption{PEC of H$_2$ calculated with TI-DMRG (solid lines) and tcDMRG (dashed lines) and cc-pVDZ (green line), cc-pVTZ (orange line), cc-pVQZ (purple line), and cc-pV5Z (violet line) basis sets.
  The bond dimension $m$ was set as discussed in the main text.
  The inset in the lower right part of the figure reports a enlarged view of the PEC around the equilibrium distance.
  The reference curve 'Ref' was taken from Ref.~\citenum{pachucki10}.}
  \label{fig:H2_CBSl}
\end{figure}

While the results reported in the previous section indicate that the basis set convergence rate is faster for tcDMRG than for TI-DMRG, it is important to verify whether the same observation holds true for molecular systems. The hydrogen molecule (H$_2$) is a good candidate for such a study, because tcDMRG calculations with large basis sets are possible and near-exact reference results are available~\cite{pachucki10}.
Therefore, we report in Figure~\ref{fig:H2_CBSl} the TI-DMRG and tcDMRG PECs of H$_2$ calculated with a sequence of basis sets cc-pVXZ.
We included all orbitals in the active space.
In each case, the bond dimension was set to twice the number of orbitals in the basis set, so that the results are essentially converged with respect to $m$.
In all H$_2$ calculations, we decreased the parameter $\gamma$ to 1/2.
We recall that $\gamma$ is the damping factor that determines the rate at which the Jastrow factor decays for large interelectronic distances.
The correlation of the only two electrons of H$_2$, each of which may be considered centred at one of the two protons, is long-ranged compared to any other many-electron molecular system.
The damping factor $\gamma$ must, therefore, be decreased to tame the decay of the Jastrow factor and to capture this long-range correlation effect.

\begin{table}[htbp!]
  \centering
  \caption{Comparison of the well depth ($D_e$, in eV) and equilibrium internuclear distance ($R_e$, in a.u.) for the hydrogen molecule obtained with TI-DMRG and tcDMRG.
  The reference results are taken from Ref.~\citenum{pachucki10}.}\vspace{0.2cm}
  \begin{tabular}{lcccc}
    \hline\hline
    basis   & \multicolumn{2}{c}{TI-DMRG} & \multicolumn{2}{c}{tcDMRG} \\
            & $D_e$ & $R_e$ & $D_e$ & $R_e$ \\
    \hline
    cc-pVDZ & 4.4936 & 1.4379 & 4.6205 & 1.4282 \\
    cc-pVTZ & 4.6998 & 1.4037 & 4.7372 & 1.4024 \\
    cc-pVQZ & 4.7322 & 1.4022 & 4.7466 & 1.4017 \\
    cc-pV5Z & 4.7411 & 1.4016 & --- & --- \\
    \hline
    reference & 4.7477 & 1.4011 & 4.7477 & 1.4011 \\
    \hline\hline
  \end{tabular}
  \label{tab:table_h2_trans}
\end{table}

We first note that tcDMRG enhances the accuracy by approximately one order of magnitude in comparison with TI-DMRG.
For example, the tcDMRG calculations in cc-pVTZ basis set are nearly as accurate as the TI-DMRG ones within the cc-pVQZ basis.
The same holds true for the tcDMRG/cc-pVQZ and TI-DMRG/cc-pV5Z pair of calculations.
A less significant improvement is observed for the double-zeta basis (cc-pVDZ), where the tcDMRG results fall short in comparison with the TI-DMRG/cc-pVTZ method.
This anomalous trend is related to the inaccuracy of the uncorrelated part of the wave function, namely the anti-symmetrized product of $1s$ orbitals on both centers.
Within the cc-pVDZ basis the $1s$ orbitals are represented by a relatively short Gaussian expansion, and the transcorrelated method is unable to improve on this description.
We expect this problem to vanish for heavier atoms, where the cc-pVDZ basis provides a more accurate description of the uncorrelated wave function.

Finally, we compare in Table~\ref{tab:table_h2_trans} two important parameters of the PEC, namely the well depth ($D_e$) and equilibrium internuclear distance ($R_e$) calculated with both methods.
We observe a similar improvement in the convergence rate with respect to the basis set size as for the raw energies.
Notably, due to cancellation of errors, the tcDMRG/cc-pVDZ results are surprisingly accurate and the well depth is comparable with TI-DMRG/cc-pVTZ result.

\subsection{Potential energy curve of N$_2$}
\label{subsec:N2}

\begin{figure}[htbp!]
  \centering
  \includegraphics[width=.9\textwidth]{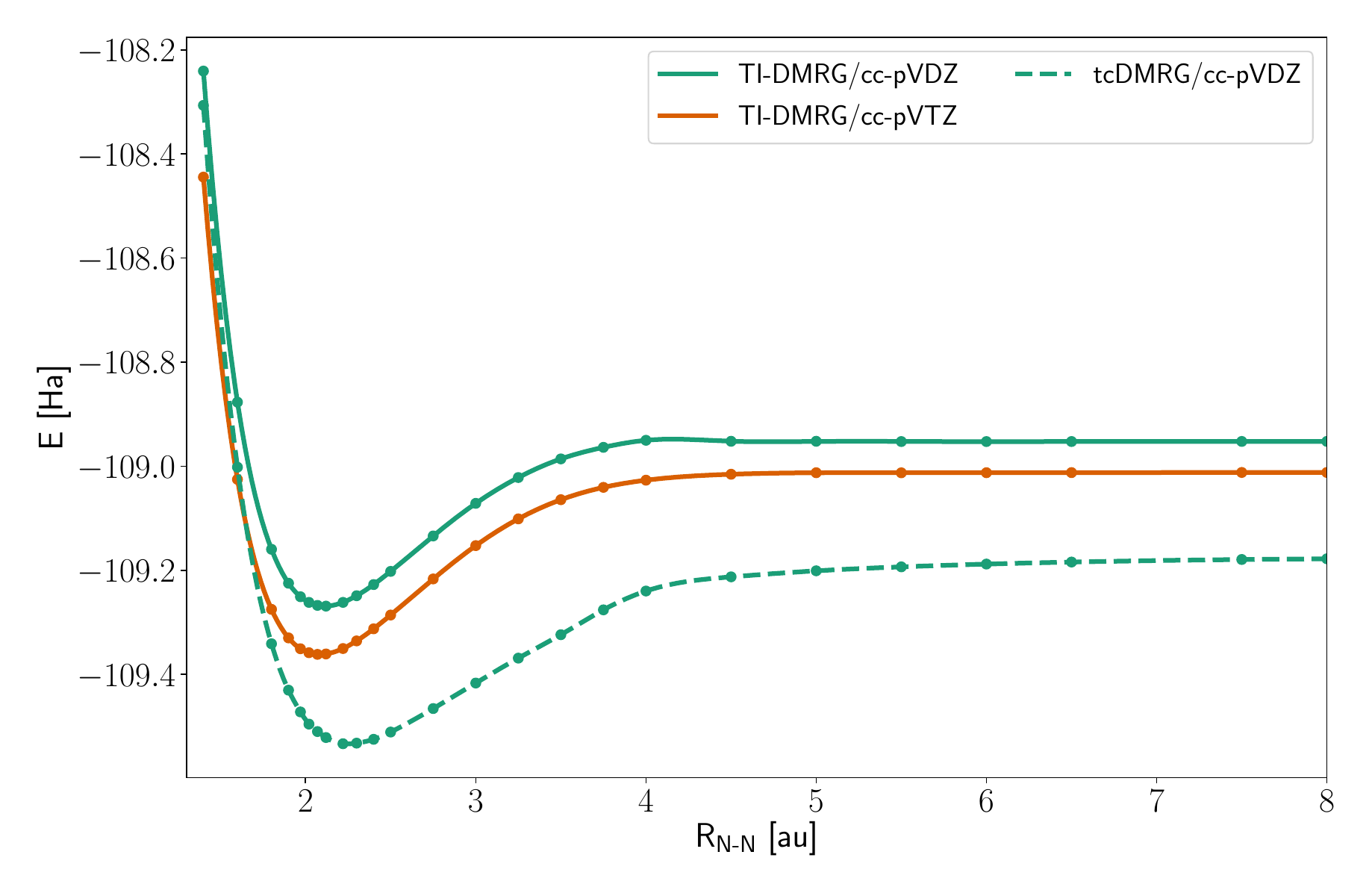}
  \caption{Potential energy curve of N$_2$ calculated with TI-DMRG (solid lines) and tcDMRG (dashed lines) with cc-pVDZ (green line) and cc-pVTZ (orange line) basis sets.
  The bond dimension $m$ is 300 for the cc-pVDZ and 450 for the cc-pVTZ basis set.}
  \label{fig:N2_CBSl}
\end{figure}

We now apply tcDMRG to calculate the PEC of the N$_2$ molecule in order to compare the method accuracy to the TI-DMRG one for many-electron systems.
We considered the $1s^2$ core orbitals of the nitrogen atoms as frozen (inactive) in the DMRG calculations, and included all remaining orbitals resulting from a given basis set in the calculations.
Hence, they become functionally equivalent to FCI.
We rely on the cc-pVDZ and cc-pVTZ basis sets for TI-DMRG, while tcDMRG calculations are limited to the cc-pVDZ basis due to memory constraints.
We fix the bond dimension at $m$=300 and $m$=450 for the cc-pVDZ and cc-pVTZ basis sets, respectively.
With these settings, in all cases the absolute energies are saturated to the level of 0.1~mHa or better.
The tcDMRG and TI-DMRG PECs are reported in Figure~\ref{fig:N2_CBSl}.
The tcDMRG/cc-pVDZ potential energy curve lies below the curves calculated with TI-DMRG and the cc-pVDZ and cc-pVTZ basis sets.
The minimum energy value is lower than the FCIQMC results obtained in Ref.~\citenum{cleland12} with a quadruple-$\zeta$ basis set.
This indicates that the tcDMRG/cc-pVDZ PEC is also lower than the FCI one in the complete basis set limit, which is a consequence of tcDMRG being a non-variational method.
Still, the convergence of the relative energies and, consequently, of the dissociation energy is faster for tcDMRG compared to TI-DMRG.
In particular, the well depth calculated with TI-DMRG is 8.61~eV and 9.51~eV within the cc-pVDZ and cc-pVTZ basis sets, respectively.
The corresponding $D_e$ value obtained with tcDMRG/cc-pVDZ calculations is 9.69~eV, while the experimental result\cite{tang05} reads 9.90~eV.
This indicates that tcDMRG accelerates the convergence of the dissociation energy of N$_2$ towards the complete basis set limit by at least one cardinal number.
We also note that both TI-DMRG/cc-pVDZ and tcDMRG/cc-pVDZ overestimate the equilibrium reference bond distance of N$_2$ calculated with TI-DMRG/cc-pVTZ.
Since the same trend is observed also for H$_2$ (see Figure~\ref{fig:H2_CBSl}), this indicates that at least a triple-zeta basis set is required to obtain accurate tcDMRG spectroscopic parameters for N$_2$.
Such a calculation is currently not feasible with our implementation of tcDMRG, which is limited to about $40-50$ active orbitals.
The main bottleneck is the storage of the MPO/MPS boundary which requires a substantial amount of memory.
Nonetheless, taking the improvements in the quality of the results into account, the tcDMRG method is competitive with TI-DMRG in terms of the computational costs.
For example, in the calculations for the N$_2$ molecule described above, the conventional DMRG/cc-pVTZ calculations required more computational time and more memory than the corresponding tcDMRG/cc-pVDZ calculations.
First, the bond dimension $m$ required to converge TI-DMRG/cc-pVTZ is higher than for tcDMRG/cc-pVDZ.
Moreover, the cost of computing the two- and three-body integrals entering the transcorrelated Hamiltonian approach is negligible in comparison with that of the DMRG optimization.

\subsection{tcDMRG for strongly correlated systems: Be$_2$}
\label{subsec:Be2}

\noindent As the last example, we study the PEC of Be$_2$ with TI-DMRG and tcDMRG with cc-pCVDZ and cc-pCVTZ basis sets to challenge the tcDMRG accuracy on a strongly correlated molecule.
Calculating the accurate dissociation Be$_2$ is a notoriously hard test-case for wave function-based methods.
On the one hand, the Be$_2$ PEC is strongly influenced by non-covalent interactions, which are described accurately only by large basis sets.
On the other hand, correlation effects are strong for Be$_2$ and single-reference methods fail in reproducing its dissociation curve.
Among the many computational studies of Be$_2$ with methods tailored for strongly-correlated systems~\cite{Bartlett1990_Be2-CC,Patkowski2007_Be2-CC,Koput2011_Be2-CC,Tenti2014_Be2-NonDynamicalCorrelation,Lesiuk2015_Be2,McKemmish2018_Be2-DFT}, the one of Sharma and co-workers~\cite{Sharma2014_Be2-DMRG} is particularly relevant in connection to this work.
In fact, also Ref.~\citenum{Sharma2014_Be2-DMRG} combines DMRG with the transcorrelated method.
However, it relies on the approach introduced by Yanai and Shiozaki~\cite{Yanai2012_CT-Transcorrelated} and constructs the transcorrelated Hamiltonian ($\mathcal{H}_\text{CT}$) through a canonical transformation in the second-quantization space:

\begin{equation}
  \mathcal{H}_\text{CT} = e^{-\mathcal{A}} \mathcal{H} e^{\mathcal{A}} \, .
  \label{eq:CanonicalTransformation}
\end{equation}

$\mathcal{A}$ is an operator expressed as a linear combination of two-particle excitation operators, where the coefficients are expressed in terms of F12-type integrals, as described in Ref.~\citenum{Yanai2012_CT-Transcorrelated}.
Unlike tcDMRG, which relies on a real-space similarity transformation, Eq.~(\ref{eq:CanonicalTransformation}) relies on a orbital space similarity transformation.
Due to this crucial difference, the Baker-Campbell-Hausdorff expansion of Eq.~(\ref{eq:CanonicalTransformation}) contains arbitrarily high many-body terms, while Eq.~(\ref{eq:SimilarityTransformedHamiltonian}) does not.
Ref.~\citenum{Sharma2014_Be2-DMRG} addresses this issue by neglecting the non-hermitian component of the two-body commutator, as well as all higher-order terms.
However, due to this approximation, the exact eigenvalues of Eq.~(\ref{eq:CanonicalTransformation}) will not coincide with those of the original, exact Hamiltonian $\mathcal{H}$.
This is not the case for tcDMRG that relies on the exact representation of Eq.~(\ref{eq:SimilarityTransformedHamiltonian}) that includes at most three-body terms.
Naturally, the price to pay is that encoding the three-body Hamiltonian in a MPO format is challenging, as we have already mentioned above.
Be$_2$ was also studied by Alavi and co-workers with transcorrelated FCIQMC,\cite{Guther2021_Be2-tcFCIQMC} who showed that the transcorrelated method enhances the convergence of FCIQMC to the complete basis set limit such that a spectroscopically accurate potential energy curve for Be$_2$ can be obtained with a triple-$\zeta$ basis set.
In this section, we verify that the same holds true for tcDMRG with the correlation factor introduced in the previous section that, as already highlighted above, differs from the one applied in the FCIQMC study.

We observed that both TI-DMRG and tcDMRG tend to converge to local minima for Be$_2$ by following the computational protocol that we applied to the other molecules (\textit{i.e.} rely on the canonical orbital ordering on the lattice and initiate the DMRG optimization from an MPS with random entries).
We adhered to the following computational procedure to enhance the convergence of both TI-DMRG and tcDMRG:

\begin{enumerate}
  \item optimize the MPS with two-site TI-DMRG, namely TI-DMRG[TS], for a given $m$ value and with the canonical orbital ordering.
  \item calculate the Fiedler ordering\cite{Barcza2011_Fiedler} based on the resulting MPS.
  \item re-optimize the MPS with TI-DMRG[TS] with the optimized Fiedler ordering.
  \item run a tcDMRG[TS,2] calculation (\textit{i.e.}, a tcDMRG[TS] calculation by neglecting the three-body part of the transcorrelated Hamiltonian) with the MPS optimized in 3. as the initial state for the imaginary-time evolution.
  \item run a tcDMRG[SS] calculation with the MPS optimized in 4. as the initial state for the imaginary-time evolution.
\end{enumerate}

\noindent Note that steps 1-4 rely on two-site DMRG algorithms.
This is a crucial requirement to avoid that the DMRG optimization converges to a local minimum.
We instead apply the single-site propagation algorithm in step 5 to tame the memory bottleneck associated with the need of storing the boundaries of the three-body operator of Eq.~(\ref{eq:TC_SecondQuantization}).
The MPS optimized by steps 1-4 will reliably approximate the exact right eigenfunction of the transcorrelated Hamiltonian.
For this reason, as we will show in the following, step~5 does not converge to local minima although we apply the single-site propagation scheme.

\begin{figure}[htbp!]
  \centering
  \includegraphics[width=\textwidth]{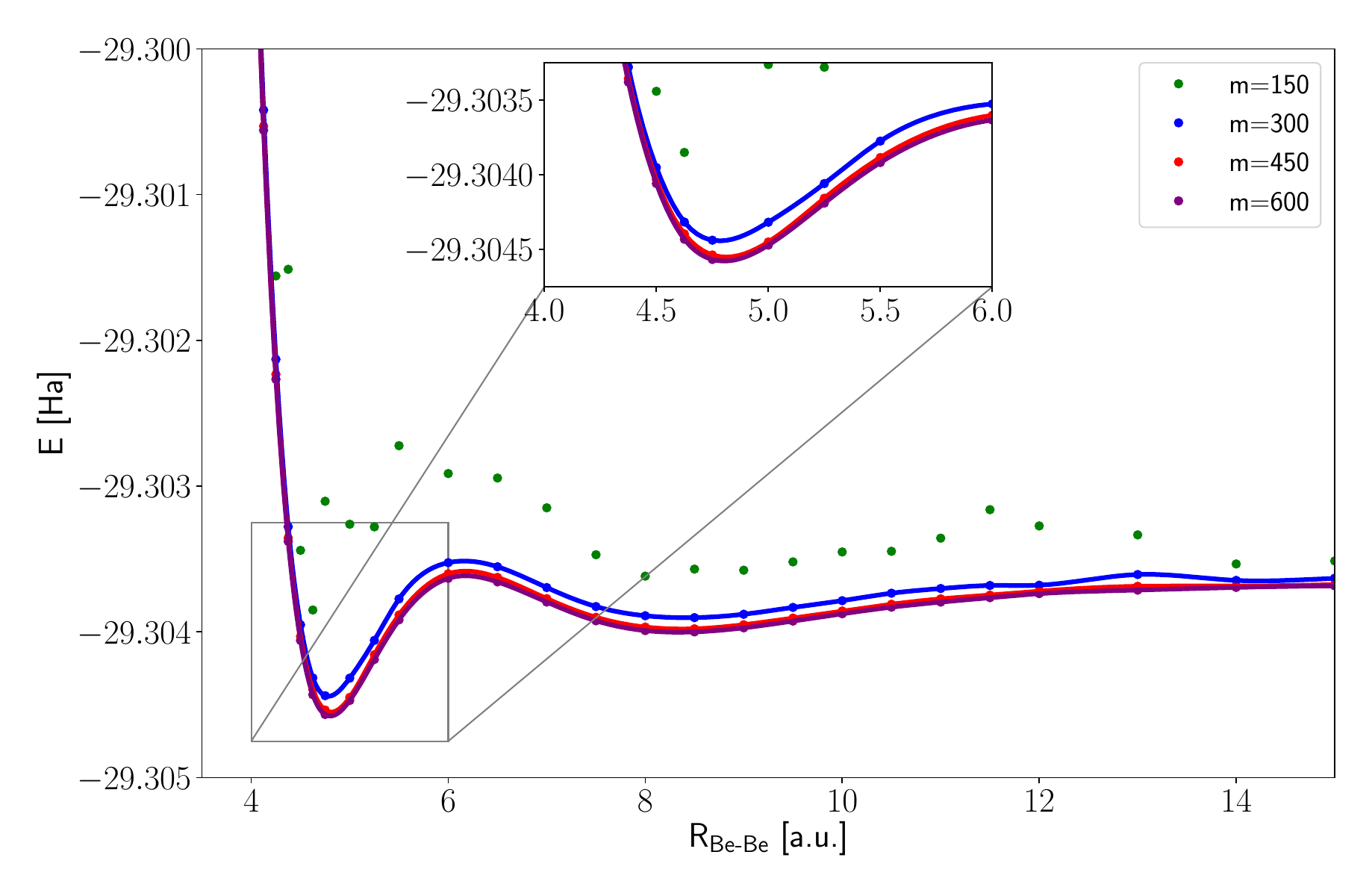}
  \caption{PEC of Be$_2$ calculated with TI-DMRG[TS] and $m$=150 (green line), $m$=300 (blue line), $m$=450 (red line), and $m$=600 (purple line).
  The basis set is cc-pCVDZ for all calculations.
  The Fiedler orbital ordering, calculated based on the MPS optimized with TI-DMRG[TS] and $m$=300, was applied to all calculations.
  The inset displays the PECs in the vicinity of the equilibrium position.
  The solid lines are obtained by fitting the electronic energies, represented as dots, with a cubic spline.}
  \label{fig:Be2_TI-DMRG_Fiedler}
\end{figure}

\noindent We report in Figure~\ref{fig:Be2_TI-DMRG_Fiedler} the PEC of Be$_2$ obtained with TI-DMRG[TS] and varying $m$ values.
We optimized in all cases the orbital ordering on the lattice with the Fiedler ordering calculated from the MPS optimized with TI-DMRG and $m$=300 (steps 2 and 3 of the computational procedure introduced above).
A physically sound PEC is obtained only starting from $m$=300, and $m$ must be further increased to 450 to obtain a quantitatively converged PEC.
No additional changes are observed by further increasing the bond dimension to 600.
However, two minima appear, one at approximately 4.75~Bohr and a second, shallower one at 9~Bohr (a result that is found independent of the chosen value for $m$).
This unphysical double-well PEC is a direct consequence of the basis set incompleteness and at least triple-$\zeta$ bases are needed in order to obtain a single-well PEC.
This has been shown both for conventional CI, in Ref.~\citenum{McKemmish2018_Be2-DFT}, and for FCIQMC, in Ref.~\citenum{Guther2021_Be2-tcFCIQMC}.
Note also that the dissociation energy calculated based on the data reported in Figure~\ref{fig:Be2_TI-DMRG_Fiedler} is 190~cm$^{-1}$, in good agreement with the FCI value of 181~cm$^{-1}$ reported in Ref.~\citenum{Patkowski2007_Be2-CC}.

\begin{figure}[htbp!]
  \centering
  \includegraphics[width=\textwidth]{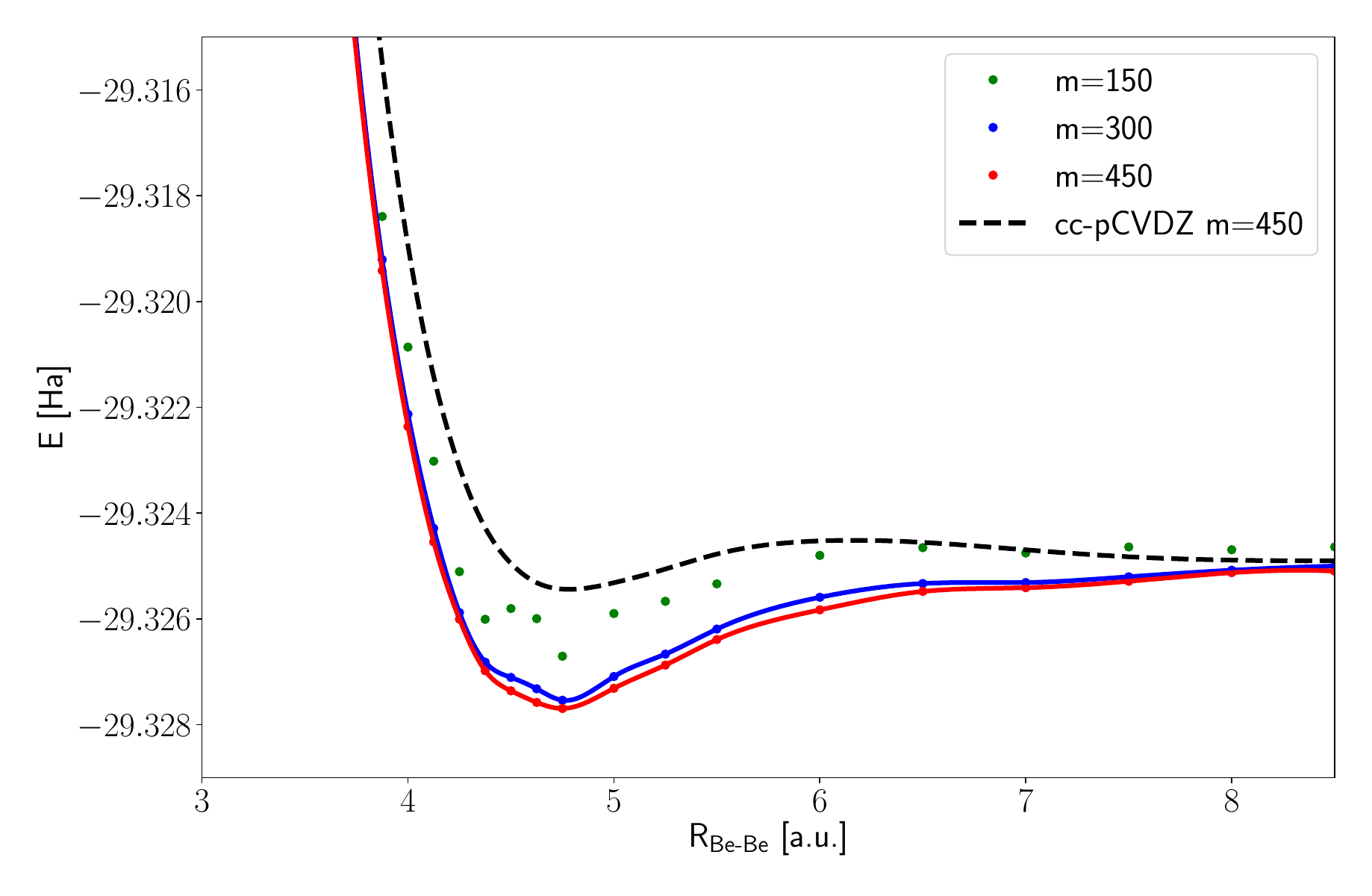}
  \caption{PEC of Be$_2$ calculated with TI-DMRG[TS], the cc-pCVTZ basis, and $m$=150 (green line), $m$=300 (blue line), and $m$=450 (red line).
  The PEC calculated with TI-DMRG[TS], the cc-pCVDZ basis, and $m$=450 is also reported (dashed, black line).
  In all cases, we mapped the orbitals onto the DMRG lattice based on the Fiedler ordering obtained from the MPS optimized with TI-DMRG[TS] and $m$=300.
  The solid lines are obtained by fitting the electronic energies, represented as dots, with a cubic spline.}
  \label{fig:Be2_TI-DMRG_Fiedler_TZ}
\end{figure}

\noindent We report in Figure~\ref{fig:Be2_TI-DMRG_Fiedler_TZ} the PEC calculated with TI-DMRG[TS] with the cc-pCVTZ basis set, together with the cc-pVDZ PEC obtained with $m$=450.
As expected, a physically acceptable PEC shape, with a single minimum at about 4.75~Bohr, is obtained with the larger triple-$\zeta$ basis set.
However, the dissociation energy (569~cm$^{-1}$ with $m$=450) still largely underestimates the experimental value\cite{Merritt2009_Be2Exp,Patkowski2009_Be2Exp} of 930~cm$^{-1}$.
To further increase the accuracy, it would be, in principle, possible to enlarge the basis set size, following the strategy of Ref.~\citenum{Sharma2014_Be2-DMRG}.
However, this would increase the lattice size and, possibly, the bond dimension $m$ required to converge the energy.
Moreover, for larger $L$ values also the risk increases that the sweep-based optimization converges to a local minimum.
Such a problem was addressed in Ref.~\citenum{Sharma2014_Be2-DMRG} by running first a DMRG calculation with a high $m$ value of 2000 and then employing the resulting MPS as a guess for lower-$m$ calculations.

\begin{figure}[htbp!]
  \centering
  \includegraphics[width=\textwidth]{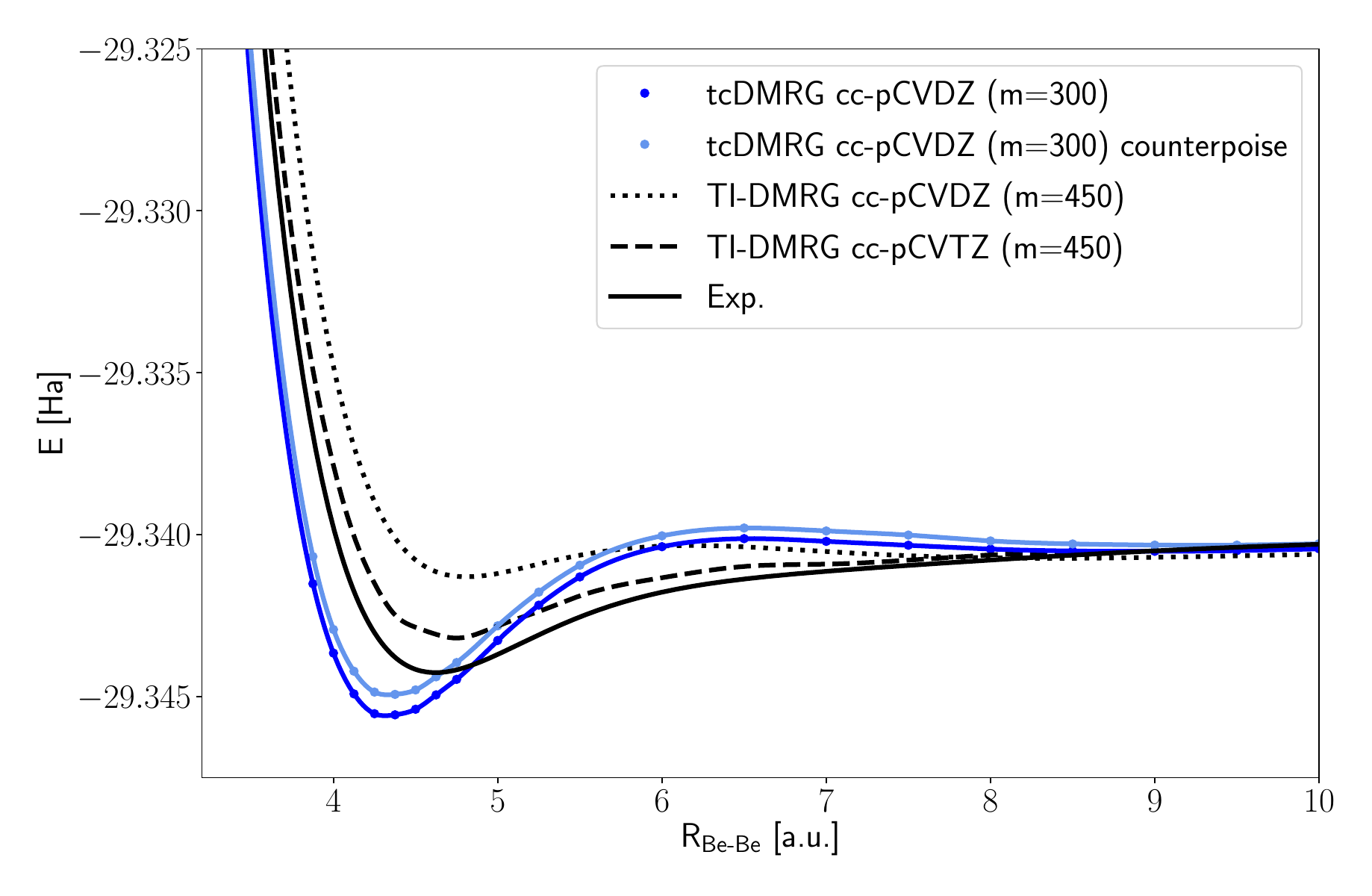}
  \caption{PEC of Be$_2$ calculated with tcDMRG, $m$=300, and the cc-pCVDZ basis set (solid, dark blue line).
  The PEC calculated with the same method and including the counterpoise correction is also reported (solid, light blue line).
  The TI-DMRG[TS] PECs calculated with $m$=400 and the cc-pCVDZ (dotted black line) and the cc-pCVTZ (dashed black line) are also reported.
  In all cases, the orbitals were sorted in the DMRG lattice based on the Fiedler ordering obtained from the MPS optimized with TI-DMRG and $m$=300.
  To facilitate the comparison between the curves, the TI-DMRG PECs calculated with the cc-pCVDZ and cc-pCVTZ basis sets have been shifted by -36.75~mHa and -15.5~mHa, respectively.
  The solid blue lines are obtained by fitting the electronic energies, represented as blue dots, with a cubic spline.
  We report also, as solid black line and labelled as ``Exp.'', the PEC taken from Ref.~\citenum{Meshov2014_Be2Curve}, which was obtained by fitting experimental spectroscopic data to an analytical potential energy function.}
  \label{fig:Be2_TC3-DMRG_Fiedler}
\end{figure}

Instead of increasing the basis set size further, we applied tcDMRG to enhance the convergence of DMRG with respect to both the basis set and the bond dimension $m$.
We report in Figure~\ref{fig:Be2_TC3-DMRG_Fiedler} the tcDMRG PEC obtained with tcDMRG by following the computational procedure introduced at the beginning of this Section.
The tcDMRG PEC obtained with the cc-pCVDZ displays a deeper minimum than the TI-DMRG[TS] PECs obtained with both the cc-pCVDZ and the cc-pCVTZ basis sets.
The tcDMRG dissociation energy is 1125~cm$^{-1}$.
The dissociation energies calculated with tcFCIQMC and a double-$\zeta$ basis set and different correlation factors are in a range between 400 and 500 cm$^{-1}$ as reported in Ref.~\citenum{Guther2021_Be2-tcFCIQMC}.
Since both FCIQMC and DMRG are efficient solvers of the FCI problem, this difference must be ascribed to the different correlation factor.
The match with the experimental value~\cite{Merritt2009_Be2Exp}, 930 cm$^{-1}$ improves compared to the TI-DMRG results, although tcDMRG still slightly overestimates the experimental dissociation energy.
By comparison with the reference curve taken from Ref.~\citenum{Meshov2014_Be2Curve}, which was calculated by fitting experimentally derived data to an analytical form of the PEC, we note that tcDMRG slightly underestimates the exact bond distance.

We further enhanced the accuracy by correcting the basis set superposition error with the well-known counterpoise correction~\cite{boys70}.
The same correction was applied in Ref.~\citenum{Guther2021_Be2-tcFCIQMC}, where the Be$_2$ PEC was studied with the transcorrelated FCIQMC method. In that work it was found that the basis set superposition error originates mostly from the $1s^2$ core electrons and that it is more effective to use a smaller basis set in the core region and apply the counterpoise correction than augment the basis with more "tight" functions.
Following this recommendation, we calculated the counterpoise correction at each $R_\text{Be-Be}$ value by subtracting from the energy of the dimer the tcDMRG energy of the Be atom calculated in the presence of a ghost atom carrying the Be cc-pCVDZ basis placed at a distance $R_\text{Be-Be}$ from the Be atom.
A bond dimension of $m$=100 is sufficient for converging these calculations for all $R_\text{Be-Be}$ values.
Figure~\ref{fig:Be2_TC3-DMRG_Fiedler} shows that the dissociation energy decreases to 1019~cm$^{-1}$ if the basis set superposition error is taken into account, which is in much better agreement with the experimental data, especially compared to the TI-DMRG results obtained with the same basis set (see Figure~\ref{fig:Be2_TI-DMRG_Fiedler}).
It is worth pointing out that the magnitude of the counterpoise correction observed in our calculations is smaller than in transcorrelated FCIQMC from Ref.~\citenum{Guther2021_Be2-tcFCIQMC}

\begin{figure}
  \centering
  \includegraphics[width=.9\textwidth]{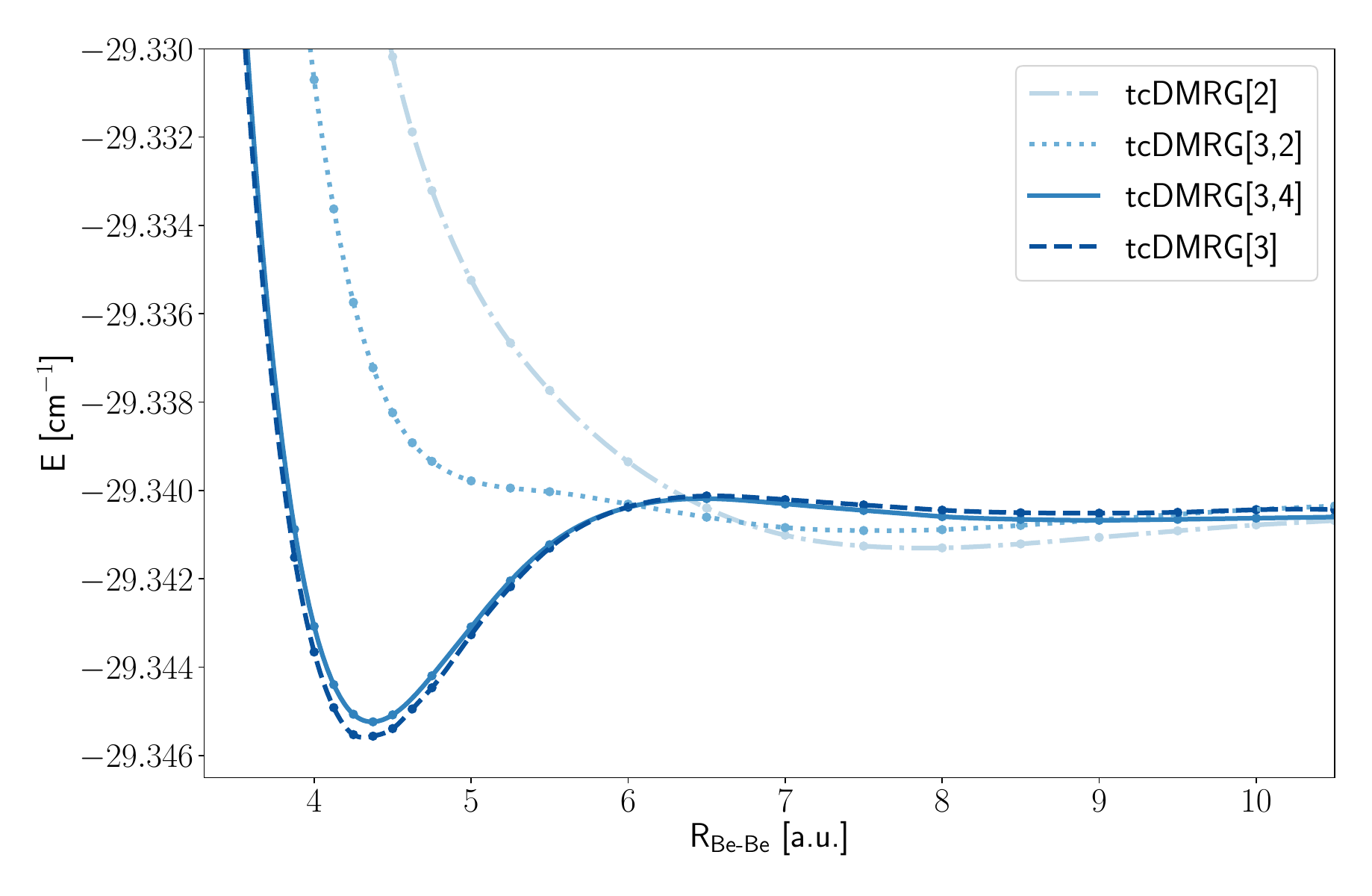}
  \caption{Be$_2$ PEC calculated with tcDMRG[2], tcDMRG[3,2], tcDMRG[3,4], and tcDMRG[3] and the cc-pCVDZ basis set.
  In all cases, we set $m$=300.
  The shade of blue increases while going from tcDMRG[2] to tcDMRG[3], \textit{i.e.} while increasing the number of terms that are included in the transcorrelated Hamiltonian.
  To facilitate the comparison, the tcDMRG[2] PEC is reported as dash-dotted line, the tcDMRG[3,2] as dotted line, the tcDMRG[3,4] PEC as solid line, and the tcDMRG[3] PEC as dashed line.
  To facilitate the comparison between the curves, the tcDMRG[2] PEC was shifted by 12.17~mHa, the tcDMRG[3,2] one by 5.37~mHa, and the tcDMRG[3,4] one by -2.17~mHa.}
  \label{fig:Be2_ManyBodyConvergence}
\end{figure}

\noindent The FCIQMC results reported in Ref.~\citenum{Guther2021_Be2-tcFCIQMC} demonstrated that a spectroscopically accurate PEC, with an accuracy of more than 10~cm$^{-1}$, can be obtained with transcorrelation and a triple-$\zeta$ basis set.
We note that, for a double-$\zeta$ basis set, our correlation factor yields more accurate results for Be$_2$ than the one employed in the FCIQMC literature reference.
This indicates that the tcDMRG method would yield spectroscopically accurate results with a triple-$\zeta$ basis set.
A complete study of the PEC of Be$_2$ will be the subject of our future studies, where we will remove the current memory bottleneck for a triple-$\zeta$ basis set by applying MPO compression algorithms. 

As we already highlighted, the main bottleneck associated with tcDMRG is that the real-space similarity transformed Hamiltonian contains the three-body term.
The dimension of exact MPO representation of these terms obtained with the algorithm presented in Refs.~\citenum{Crosswhite2008_FiniteAutomata,Keller2015_MPS-MPO-QuantumChemical} scales as $\mathcal{O}(L^5)$, compared to the $\mathcal{O}(L^3)$ scaling of the original TI-DMRG method.
Currently, this limits tcDMRG calculations to lattices with up to (approximately) 40 sites, whereas TI-DMRG are feasible for up to $L$=100.
It would be desirable to find approximations to Eq.~(\ref{eq:SimilarityTransformedHamiltonian}) that can be encoded as more compact MPOs.
Methods relying on canonical transformation theory truncate Eq.~(\ref{eq:CanonicalTransformation}) by approximating the two-body potential and by neglecting any three- or higher-order terms.
Similarly, it has been recently shown for the transcorrelated CC method\cite{Schraivogel2021_TCC-Be2} that Eq.~(\ref{eq:CanonicalTransformation}) can be efficiently approximated by its two-body normal ordered form.
A thorough analysis of strategies to compress the MPO representation of the transcorrelated Hamiltonian is beyond the scope of the present work.
Here, we analyze the efficiency of a very simple approximation.
We introduce the tcDMRG[3,$i$] method that approximates the three-body term of the transcorrelated Hamiltonian by including only strings of second-quantized operators coupling up to $i$ different sites.
For instance, tcDMRG[3,4] neglects all strings of second-quantized operators where 5 and 6 different indices appear and, therefore, can be considered a rough approximation of the normal-ordered two-body transcorrelated Hamiltonian.
The MPO complexity of tcDMRG[3,4] is the same as in conventional TI-DMRG.
As we show in Figure~\ref{fig:Be2_ManyBodyConvergence}, the tcDMRG[3] and tcDMRG[3,4] curves are nearly identical (note that the tcDMRG[3,4] PEC was shifted by 2.17~mHa towards higher energy values), and the dissociation energy is 1053~cm$^{-1}$, in very good agreement with the tcDMRG[3] result of 1125~cm$^{-1}$.
This result agrees with the findings reported in Ref.~\citenum{Sharma2014_Be2-DMRG}, where accurate results were obtained with a two-body approximation of the canonical transformation Hamiltonian.
A key advantage of tcDMRG is that the terms that are neglected are known exactly.
Their effect could be added \textit{a posteriori} through, for instance, perturbation theory.
We will investigate this possibility in future work.

\section{Conclusions}

In this work, we introduced the transcorrelated DMRG algorithm for the full electronic Hamiltonian in
second quantized form to deal with dynamical correlations induced by the electron-electron cusp. 
We addressed the problem associated with the lack of hermiticity of the transcorrelated Hamiltonian by encoding it 
as a matrix product operator\cite{Keller2015_MPS-MPO-QuantumChemical} and by optimizing the right 
eigenfunction associated with the ground state with imaginary-time time-dependent DMRG.\cite{Baiardi2020_tcDMRG}
We demonstrated how the density fitting approximation can be leveraged to calculate efficiently the two- and three-body 
integrals entering the second quantization representation of the transcorrelated Hamiltonian.

We then applied tcDMRG to calculate ground-state electronic energies of atoms and first-row diatomic molecules.
We showed that, compared to conventional DMRG, tcDMRG convergence to the complete basis set limit is significantly enhanced.
As has already been shown for FCIQMC,\cite{Alavi2019_tcFCIQMC-Molecules} the transcorrelation 
similarity transformation reduces the many-body wave function entanglement, which makes the wave function
easier to represent as a matrix product state. This is akin to the effect that we observed for a
 short-range density-functional-based representation of the electron-electron interaction\cite{Hedegard2015},
which also addresses the electron-electron cusp problem.

A limitation of tcDMRG is that the three-body transcorrelated Hamiltonian is encoded as a highly non-compact matrix product operator.
This significantly increases the computational cost associated with the tensor contractions that are evaluated 
in DMRG iteration steps.
However, we showed for Be$_2$ that the Hamiltonian terms that couple five or six different sites of the DMRG lattice have only a minor effect on the tcDMRG accuracy.
This suggests that the MPO representation of the transcorrelated Hamiltonian can be largely compressed without compromising accuracy, which makes tcDMRG amenable for specific MPO compression techniques \cite{Hubig2017_MPO,Zalatel2020_LocalMPO}.

As the results reported in this work suggest that tcDMRG can enhance the accuracy of conventional DMRG calculations,
we are encouraged to further expand on the tcDMRG approach. For instance,
a spin-adapted form of the tcDMRG method may be obtained by relying on the framework defined in Ref.~\citenum{Keller2016_SpinAdapted} and can further enhance the efficiency of tcDMRG, especially for electronically excited states.
Moreover, by reducing the wave function entanglement, tcDMRG makes it easier to distinguish weakly- and strongly-correlated orbitals, which is beneficial for fully automated active space selection\cite{Stein2016a,Stein2016b,Stein2017,Stein2019}.
This will also allow one to apply tcDMRG effectively to large strongly-correlated molecular systems.

Surely, missing (long-range) dynamical correlation effects must be taken into account for accurate
electronic energies. They can be considered with standard methods such as those based on perturbation theory
(see, for instance, Refs.\ \citenum{Kurashige2011_CASPT2,Yanai2013_DMRG-CASPT2,Freitag2017_NEVPT2}) or on 
coupled cluster variants\cite{Veis2016_TCC,Morchen2020_TCC,Lee2021_ExternallyCorrectedCC-DMRG,Guther2021_Be2-tcFCIQMC}.

\section*{Appendix: Derivation of Eq. (40)}
Here, we provide a detailed derivation of the key equation Eq. (40). 
Note that essentially the same derivation follows for Eqs. (41) and (42). 
As before, we assume that all orbitals are purely real.
First, we rewrite Eq. (40) explicitly without the bracket notation
\begin{align}
\begin{split}
    \alpha_{PQ}^{\kappa\tau} &= (P|\nabla_1 f(r_{12} )|Q|\nabla_1 f(r_{13})|\kappa\tau) = \\
    &=\iiint d\textbf{r}_1\,d\textbf{r}_2\,d\textbf{r}_3\,
    \chi_P(\textbf{r}_1)\,\nabla_1 f(r_{12})\,\chi_Q(\textbf{r}_2)\,\nabla_1 f(r_{13})\,
    \varphi_\kappa(\textbf{r}_3)\,\varphi_\tau(\textbf{r}_3).
\end{split}
\end{align}
In the above formula we artificially introduce integration over the fourth electron and insert the Dirac delta function $\delta(\textbf{r}_1-\textbf{r}_4)$ as follows
\begin{align}
\begin{split}
    \alpha_{PQ}^{\kappa\tau} = 
    -\iiiint d\textbf{r}_1\,d\textbf{r}_2\,d\textbf{r}_3\,d\textbf{r}_4\,
    \chi_P(\textbf{r}_1)\,\nabla_1 f(r_{12})\,\chi_Q(\textbf{r}_2)\,\nabla_3 f(r_{34})\,
    \varphi_\kappa(\textbf{r}_3)\,\varphi_\tau(\textbf{r}_3)\,\delta(\textbf{r}_1-\textbf{r}_4),
\end{split}
\end{align}
where we have additionally exploited an elementary identity $\nabla_4 f(r_{43})=-\nabla_3 f(r_{43})$.
Next, in order to separate the integration over the coordinates of electron pairs $12$ and $34$, resolution of identity is used
\begin{align}
    \delta(\textbf{r}_1-\textbf{r}_4) = \phi_x(\textbf{r}_1)\,\phi_x(\textbf{r}_4),
\end{align}
where $\phi_x$ are the elements of the orthonormal RI basis. This leads to
\begin{align}
\begin{split}
    \alpha_{PQ}^{\kappa\tau} &= 
    -\iint d\textbf{r}_1\,d\textbf{r}_2\,
    \chi_P(\textbf{r}_1)\,\phi_x(\textbf{r}_1)\,\nabla_1 f(r_{12})\,\chi_Q(\textbf{r}_2) \\
    &\times \iint d\textbf{r}_3\,d\textbf{r}_4\, 
    \varphi_\kappa(\textbf{r}_3)\,\varphi_\tau(\textbf{r}_3)\,
    \nabla_3 f(r_{34})\,\phi_x(\textbf{r}_4).
\end{split}
\end{align}
In the final step, we use the divergence theorem to transfer the $\nabla$ operators from the correlation factors to the functions in the brackets. This is dictated mostly by simplifications in the implementation, as the basis set functions (which are Gaussian functions in the present work) are easy to differentiate. We arrive at
\begin{align}
\begin{split}
    \alpha_{PQ}^{\kappa\tau} &= -
    \iint d\textbf{r}_1\,d\textbf{r}_2\,
    \nabla_1\Big[\chi_P(\textbf{r}_1)\,\phi_x(\textbf{r}_1)\Big]\,
    f(r_{12})\,\chi_Q(\textbf{r}_2) \\
    &\times \iint d\textbf{r}_3\,d\textbf{r}_4\, 
    \nabla_3\Big[\varphi_\kappa(\textbf{r}_3)\,\varphi_\tau(\textbf{r}_3)\Big]\,
    f(r_{34})\,\phi_x(\textbf{r}_4).
\end{split}
\end{align}
If we now return to the bracket notation, we can rewrite the above formula as
\begin{align}
    \alpha_{PQ}^{\kappa\tau} &= -\big(\nabla[Px]\big|f(r_{12})\big|Q\big)\, 
    \big(\nabla[\kappa\tau]\big|f(r_{12})\big|x\big),
\end{align}
which is Eq. (40).

\section*{Acknowledgements}
ML acknowledges the support by the Polish National Agency of Academic Exchange through the Bekker programme No. PPN/BEK/2019/1/00315/U/00001.

%\bibliography{biblio}

\providecommand{\latin}[1]{#1}
\makeatletter
\providecommand{\doi}
  {\begingroup\let\do\@makeother\dospecials
  \catcode`\{=1 \catcode`\}=2 \doi@aux}
\providecommand{\doi@aux}[1]{\endgroup\texttt{#1}}
\makeatother
\providecommand*\mcitethebibliography{\thebibliography}
\csname @ifundefined\endcsname{endmcitethebibliography}
  {\let\endmcitethebibliography\endthebibliography}{}

\end{document}